\documentclass[12pt]{article}
\usepackage{amssymb}
\usepackage{amsmath}
\usepackage{graphicx}
\usepackage{indentfirst}
\usepackage{cite}

\allowdisplaybreaks

\begin{document}

\title{Cross Sections for Inelastic 2-to-2 Meson-Meson 
\\ Scattering in Hadronic Matter}
\author{Ting-Ting Wang and Xiao-Ming Xu}
\date{}
\maketitle \vspace{-1cm}
\centerline{Department of Physics, Shanghai University, Baoshan,
Shanghai 200444, China}

\begin{abstract}
With quark-antiquark annihilation and creation in the first Born approximation,
we study the reactions: $K \bar {K} \to K \bar {K}^\ast, 
~K \bar{K} \to K^* \bar{K}, ~\pi K \to \pi K^\ast, ~\pi K \to \rho K,
~\pi \pi \to K \bar{K}^\ast, ~\pi \pi \to K^\ast \bar{K},
~\pi \pi \to K^\ast \bar{K}^\ast, ~\pi \rho \to K \bar{K},
~\pi \rho \to K^\ast \bar{K}^\ast, ~\rho \rho \to K^\ast \bar{K}^\ast,
~K \bar{K}^\ast \to \rho \rho$, and $K^* \bar{K} \to \rho \rho$. 
Unpolarized cross sections for the reactions are obtained from transition
amplitudes that are composed of mesonic quark-antiquark relative-motion
wave functions and the transition potential for quark-antiquark annihilation
and creation. From a quark-antiquark potential that is equivalent to the 
transition potential, we prove that the total spin of the two final mesons
may not equal the total spin of the two initial mesons. Based on flavor
matrix elements, cross sections for some isospin channels of reactions can be
obtained from the other isospin channels of reactions. Remarkable temperature
dependence of the cross sections is found.
\end{abstract}



\vspace{0.5cm}
\leftline{\bf I. INTRODUCTION}
\vspace{0.5cm}

In hadronic matter that is created in ultrarelativistic heavy-ion collisions,
various kinds of meson-meson scattering take place. The meson-meson scattering
can be studied in quark degrees of freedom or meson degrees of freedom.
Elastic $\pi \pi$ scattering for $I=2$ and elastic $\pi K$ scattering for
$I=3/2$ have been studied in the quark interchange mechanism in the first
Born approximation \cite{BS} and in
nonperturbative schemes together with chiral perturbation theory
\cite{CGL,DHT,DP,GPSCZ}.
The reactions $\pi \pi \to \rho \rho$ for $I=2$, $KK \to K^*K^*$ for $I=1$,
$\pi K \to \rho K^*$ for $I=3/2$, and so on have also been studied in the quark
interchange mechanism \cite{LX,SX}. 
The reactions $\pi \pi \to K
\bar K$, $\rho \rho \to K \bar K$, $\pi \rho \to K \bar {K}^*$, and
$\pi \rho \to K^* \bar {K}$ can be studied by quark-antiquark
annihilation and creation in the first Born approximation \cite{SXW} or
through one-meson exchange
in effective meson Lagrangians \cite{LDHS,BKWX}. Furthermore, the four
isospin channels, $\pi K \to \rho K^\ast$ for $I=1/2$, $\pi K^\ast \to \rho K$
for $I=1/2$, $\pi K^\ast \to \rho K^\ast$ for $I=1/2$, and
$\rho K \to \rho K^\ast$ for $I=1/2$, are studied in the assumption that quark
interchange as well as quark-antiquark annihilation and creation are dominant
mechanisms \cite{YXW}. These studies have revealed interesting features of 
these reactions. For example, the cross
section for the inelastic meson-meson scattering governed by quark interchange
increases very rapidly to a maximum value and then decreases rapidly while the
center-of-mass energy of the two colliding mesons increases from threshold; 
however, the cross section for the inelastic meson-meson scattering governed by
quark-antiquark annihilation and creation may decrease very slowly from the
maximum value.

In the present work we are interested in the inelastic meson-meson scattering
among $\pi$, $\rho$, $K$, and $K^\ast$ mesons, which is assumed to be
dominated
by quark-antiquark annihilation and creation. The meson-meson reactions studied
in Ref. \cite{SXW} include $\pi \pi \to \rho \rho$, $K \bar
{K} \to K^* \bar {K}^\ast$, $K \bar{K}^\ast \to K^* \bar{K}^\ast$, $K^\ast
\bar{K} \to K^* \bar{K}^\ast$, $\pi \pi \to K \bar K$, $\pi \rho \to
K \bar {K}^\ast$, $\pi \rho \to K^* \bar{K}$, and $K \bar {K} \to \rho \rho$.
But these reactions do not exhaust all the 2-to-2 meson-meson reactions among
$\pi$, $\rho$, $K$, and $K^\ast$ mesons. Hence,
in the present work we study these reactions:
$K \bar {K} \to K \bar {K}^\ast,~ K \bar{K} \to K^* \bar{K},
~\pi K \to \pi K^\ast,~ \pi K \to \rho K,
~\pi \pi \to K \bar{K}^\ast,~ \pi \pi \to K^\ast \bar{K},
~\pi \pi \to K^\ast \bar{K}^\ast, ~\pi \rho \to K \bar{K},
~\pi \rho \to K^\ast \bar{K}^\ast, ~\rho \rho \to K^\ast \bar{K}^\ast,
~K \bar{K}^\ast \to \rho \rho$, and $K^* \bar{K} \to \rho \rho$.
These reactions have not been studied elsewhere, and
complement those reactions studied in Ref. \cite{SXW}.

This paper is organized as follows. In Sec.~II we present formulas of
unpolarized cross sections for 2-to-2 meson-meson reactions that are governed
by annihilation of a quark-antiquark pair and creation of 
another quark-antiquark pair. In Sec.~III we calculate transition amplitudes
with mesonic quark-antiquark wave functions and
the transition potential for quark-antiquark annihilation and creation.
In Sec.~IV we show unpolarized cross sections for 
inelastic meson-meson scattering, and give relevant discussions. In Sec.~V we
summarize the present work.

\vspace{0.5cm}
\leftline{\bf II. CROSS-SECTION FORMULAS}
\vspace{0.5cm}

It is shown by the two Feynman diagrams in Fig. 1 that the reaction 
$A+B \to C+D$ is caused by
quark-antiquark annihilation and creation in the Born approximation.
In the left diagram of Fig. 1 the quark of meson $A$ and the antiquark of meson
$B$ annihilate into a gluon, this gluon creates a new quark-antiquark pair, 
and the new quark and the antiquark in meson $A$ combine into 
meson $C$ as well as the new antiquark and the quark in meson $B$ combine into
meson $D$. In the right diagram of Fig. 1 the antiquark of meson $A$ and the
quark of meson $B$ annihilate into a gluon, the gluon creates a new
quark-qnatiquark pair, and the new antiquark and the quark of
meson $A$ form meson $C$ as well as the new quark and the antiquark of meson
$B$ form meson $D$. These are the processes that we consider in the present 
work.

\begin{figure}[htbp]
  \centering
    \includegraphics[scale=0.8]{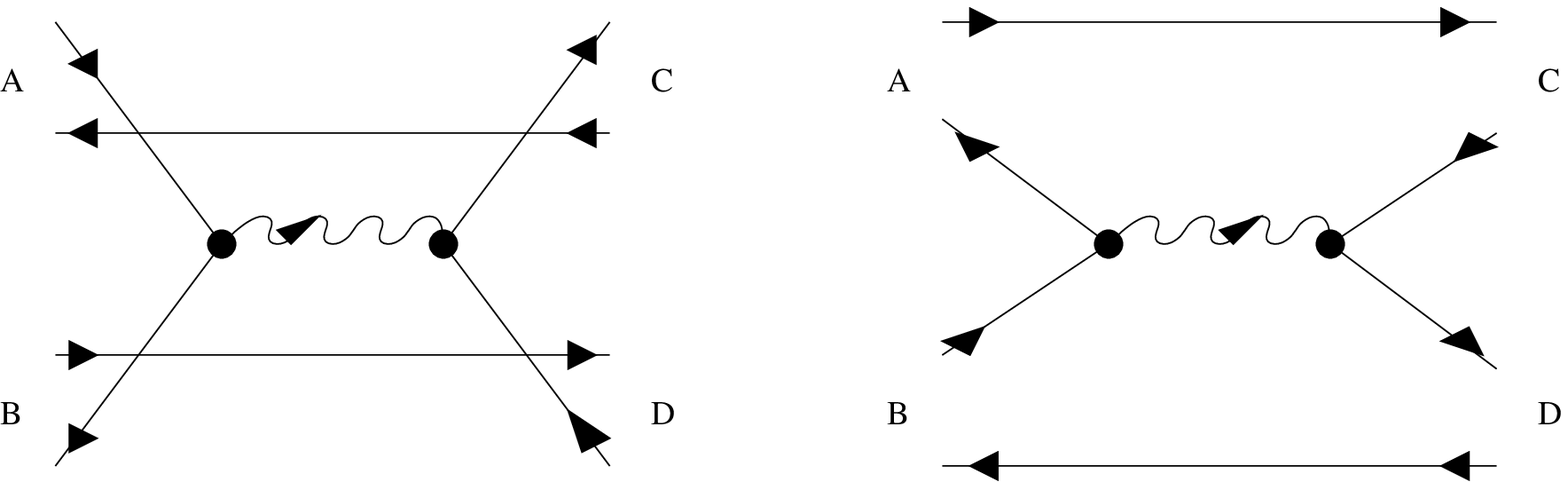}
\caption{Solid (wavy) lines stand for quarks or antiquarks (gluon).
The left diagram has $q_{1}+\bar{q}_2 \to q_{3}+\bar{q}_4$ for 
$A(q_1\bar{q}_1)+B(q_2\bar{q}_2) \to C(q_3\bar{q}_1)+D(q_2\bar{q}_4)$,
and the right diagram has $\bar{q}_1 + q_2 \to q_3 + \bar{q}_4$ for 
$A(q_1\bar{q}_1)+B(q_2\bar{q}_2) \to C(q_1\bar{q}_4)+D(q_3\bar{q}_2)$.}
\label{fig1}
\end{figure}

Let $E_A$ ($E_B$, $E_C$, $E_D$) and $J_{Az}$ ($J_{Bz}$, $J_{Cz}$, $J_{Dz}$) 
denote the energy and the magnetic projection quantum number of the angular 
momentum $J_A$ ($J_B$, $J_C$, $J_D$) of meson $A$ $( B, C, D)$, 
respectively. From the four-momenta of mesons $A$ and $B$, $P_A$ and $P_B$,
the Mandelstam variable $s=(P_A+P_B)^2$ is defined. The unpolarized cross 
section for $A+B \to C+D$ depends on $\sqrt s$ and temperature $T$. Let
$\theta$ be the angle between $\vec{P}$ and $\vec{P}'$ which are the 
three-dimensional momenta of mesons $A$ and $C$ in the center-of-mass frame, 
respectively. The unpolarized cross section is \cite{SXW}
\begin{eqnarray}
\sigma^{\rm unpol}(\sqrt {s},T) & = & \frac {1}{(2J_A+1)(2J_B+1)}
\frac{1}{32\pi s}\frac{|\vec{P}^{\prime }(\sqrt{s})|}{|\vec{P}(\sqrt{s})|}
              \nonumber    \\
&\times & \int_{0}^{\pi }d\theta \sum\limits_{J_{Az}J_{Bz}J_{Cz}J_{Dz}}
\mid {\cal M}_{{\rm a}q_1\bar{q}_2}+{\cal M}_{{\rm a}\bar{q}_1q_2} \mid^2
\sin \theta ,
\label{eq1}
\end{eqnarray}
where  ${\cal M}_{{\rm a}q_1\bar{q}_2}$ and
${\cal M}_{{\rm a}\bar{q}_1q_2}$ are the transition amplitudes corresponding 
to the left diagram and the right diagram in Fig.~\ref{fig1}, respectively. 
The transition amplitudes are given by
\begin{eqnarray}
{\cal M}_{{\rm a}q_1\bar {q}_2} & = & 
\frac {(m_{q_3}+m_{\bar{q}_1})^3}{m_{\bar{q}_1}^3}
\sqrt {2E_A2E_B2E_C2E_D}
\int d\vec{r}_{q_1\bar{q}_1} d\vec{r}_{q_2\bar{q}_4} 
d\vec{r}_{q_3\bar{q}_1,q_2\bar{q}_4} 
        \nonumber   \\
& &
\psi_{q_3\bar {q}_1}^+ (\vec {r}_{q_3\bar {q}_1})
\psi_{q_2\bar {q}_4}^+ (\vec {r}_{q_2\bar {q}_4})
V_{{\rm a}q_1\bar{q}_2}
\psi_{q_1\bar {q}_1} (\vec {r}_{q_1\bar {q}_1})
\psi_{q_2\bar {q}_2} (\vec {r}_{q_2\bar {q}_2})
       \nonumber   \\
& &
e^{i\vec {p}_{q_1\bar {q}_1,q_2\bar {q}_2}
\cdot \vec {r}_{q_1\bar {q}_1,q_2\bar {q}_2}
-i\vec {p}_{q_3\bar {q}_1,q_2\bar {q}_4}
\cdot \vec {r}_{q_3\bar {q}_1,q_2\bar {q}_4}},
\label{eq2}
\end{eqnarray}
\begin{eqnarray}
{\cal M}_{{\rm a}\bar{q}_1q_2} & = & 
\frac {(m_{q_1}+m_{\bar{q}_4})^3}{m_{q_1}^3}
\sqrt {2E_A2E_B2E_C2E_D}
\int d\vec{r}_{q_1\bar{q}_1} d\vec{r}_{q_3\bar{q}_2} 
d\vec{r}_{q_1\bar{q}_4,q_3\bar{q}_2} 
        \nonumber   \\
& &
\psi_{q_1\bar {q}_4}^+ (\vec {r}_{q_1\bar {q}_4})
\psi_{q_3\bar {q}_2}^+ (\vec {r}_{q_3\bar {q}_2})
V_{{\rm a}\bar{q}_1q_2}
\psi_{q_1\bar {q}_1} (\vec {r}_{q_1\bar {q}_1})
\psi_{q_2\bar {q}_2} (\vec {r}_{q_2\bar {q}_2})
       \nonumber   \\
& &
e^{i\vec {p}_{q_1\bar {q}_1,q_2\bar {q}_2}
\cdot \vec {r}_{q_1\bar {q}_1,q_2\bar {q}_2}
-i\vec {p}_{q_1\bar {q}_4,q_3\bar {q}_2}
\cdot \vec {r}_{q_1\bar {q}_4,q_3\bar {q}_2}},
\label{eq3}
\end{eqnarray}
where $m_a$ is the mass of constituent $a$;
$V_{{\rm a}q_1\bar{q}_2}$ and $V_{{\rm a}\bar{q}_1q_2}$ are 
the transition potentials for $q_1+\bar{q}_2 \to q_3+\bar{q}_4$
in the left diagram and $\bar{q}_1+q_2 \to q_3+\bar{q}_4$ in the right 
diagram of Fig.~\ref{fig1}, respectively; 
$\psi_{ab}$ and $\vec{r}_{ab}$ are the wave function and the relative
coordinate of constituents $a$ and $b$, respectively.
The relative coordinate and the relative momentum of $q_1\bar {q}_1$ and 
$q_2\bar {q}_2$ are denoted by $\vec {r}_{q_1\bar {q}_1,q_2\bar {q}_2}$ and
$\vec {p}_{q_1\bar {q}_1,q_2\bar {q}_2}$, respectively;
similar meanings apply to $\vec {r}_{q_3\bar {q}_1,q_2\bar {q}_4}$,
$\vec {p}_{q_3\bar {q}_1,q_2\bar {q}_4}$, 
$\vec {r}_{q_1\bar {q}_4,q_3\bar {q}_2}$, and
$\vec {p}_{q_1\bar {q}_4,q_3\bar {q}_2}$.

\vspace{0.5cm}
\leftline{\bf III. TRANSITION AMPLITUDE}
\vspace{0.5cm}

In order to calculate the transition amplitudes, we need the mesonic
quark-antiquark wave functions and the transition 
potential for quark-antiquark annihilation and creation. 
The quark-antiquark annihilation and creation is shown in Fig. 2 with 
$q (p_1) + \bar{q} (-p_2) \to q^{~\prime} (p_3) + \bar{q}^{~\prime} (-p_4)$, 
where $p_1$ and $p_3$ ($p_2$ and $p_4$) are the four-momenta of quarks 
(antiquarks).
The transition potential shown below is given in Ref. \cite{SXW},
\begin{eqnarray}
V_{{\rm a}q\bar{q}}(\vec {k}) & = &
\frac {g_{\rm s}^2}{k^2}\frac {\vec {\lambda}(34)}{2} \cdot
\frac {\vec {\lambda}(21)}{2} \left(
\frac {\vec {\sigma}(34) \cdot \vec{k} \vec {\sigma}(21) \cdot \vec{k}}
{4m_{q^\prime} m_q} -\vec {\sigma}(34) \cdot \vec {\sigma}(21)  \right.
                       \nonumber     \\
& - &
\left. \frac {\vec {\sigma}(21) \cdot \vec{p}_2 \vec {\sigma}(34)
\cdot \vec {\sigma}(21) \vec {\sigma}(21) \cdot \vec{p}_1}{4m_q^2}
-\frac {\vec {\sigma}(34) \cdot \vec{p}_3 \vec {\sigma}(34)
\cdot \vec {\sigma}(21) \vec {\sigma}(34) \cdot \vec{p}_4}
{4m_{q^\prime}^2} \right), ~~~~~~
\end{eqnarray}
where $\vec{k}$ is the three-dimensional momentum of gluon, 
$g_s$ is the gauge coupling constant, $m_{q}$ ($m_{q'}$) is the mass of 
the initial (final) quark, $\vec \lambda$ are the Gell-Mann matrices, and
$\vec \sigma$ are the Pauli matrices. 
$\vec{\lambda}(21)$ ($\vec{\sigma}(21)$) mean that they have matrix elements 
between the color (spin) wave functions of the initial quark and the initial
antiquark, and $\vec{\lambda}(34)$ ($\vec {\sigma}(34)$) mean that they have
matrix elements between the color (spin) wave functions of the final quark and
the final antiquark.

The wave functions of mesons $A$ , $B$, $C$, and $D$ are individually given by
\begin{equation}
\psi_A =\phi_{A\rm rel} \phi_{A\rm color} \phi_{A\rm flavor} \chi_{S_A S_{Az}},
\label{eq6}
\end{equation}
\begin{equation}
\psi_B =\phi_{B\rm rel} \phi_{B\rm color} \phi_{B\rm flavor} \chi_{S_B S_{Bz}},
\label{eq7}
\end{equation}
\begin{equation}
\psi_C =\phi_{C\rm rel} \phi_{C\rm color} \phi_{C\rm flavor} \chi_{S_C S_{Cz}},
\label{eq8}
\end{equation}
\begin{equation}
\psi_D =\phi_{D\rm rel} \phi_{D\rm color} \phi_{D\rm flavor} \chi_{S_D S_{Dz}},
\label{eq9}
\end{equation}
where $S_{i}$ is the spin of meson $i$, and $S_{iz}$ is its magnetic projection
quantum number. According to the two diagrams in Fig. 1, 
$\psi_A = \psi_{q_1\bar{q}_1}$, $\psi_B = \psi_{q_2\bar{q}_2}$, 
$\psi_C = \psi_{q_3\bar{q}_1} = \psi_{q_1\bar{q}_4}$, and
$\psi_D = \psi_{q_2\bar{q}_4} = \psi_{q_3\bar{q}_2}$. The wave function of 
meson $i$ is made up of the quark-antiquark relative-motion wave
function $\phi_{i\rm rel}$, the color wave function $\phi_{i\rm color}$, 
the flavor wave function $\phi_{i\rm flavor}$, and the spin wave function 
$\chi_{S_i S_{iz}}$.

The transition amplitudes contain color, spin, and flavor matrix elements.
The color matrix elements related to the Gell-Mann matrices are 4/9 for the two
diagrams in Fig. 1. The spin matrix elements related to the Pauli matrices have
been provided in Ref. \cite{SXW}. Let ${\cal M}_{{\rm a}q_1\bar{q}_2\rm f}$
and ${\cal M}_{{\rm a}\bar{q}_1q_2\rm f}$ represent the flavor matrix elements
that correspond to the left and right diagrams in Fig. 1, respectively.
The values of the flavor matrix elements are listed in Table~\ref{table1}, 
where $I$ is the total isospin of the two initial or final mesons for the 
reactions:
\begin{eqnarray}
\nonumber
K \bar {K} \to K \bar {K}^\ast, ~K \bar{K} \to K^* \bar{K},
~\pi K \to \pi K^\ast, ~\pi K \to \rho K,
~\pi \pi \to K \bar{K}^\ast, ~\pi \pi \to K^\ast \bar{K},
\\  \nonumber
\pi \pi \to K^\ast \bar{K}^\ast, ~\pi \rho \to K \bar{K},
~\pi \rho \to K^\ast \bar{K}^\ast, ~\rho \rho \to K^\ast \bar{K}^\ast, 
~K \bar{K}^\ast \to \rho \rho, ~K^* \bar{K} \to \rho \rho.
\end{eqnarray}

The quark-antiquark relative-motion wave functions are given by
the Schr\"odinger equation with a temperature-dependent potential.
The experimental masses of ground-state mesons \cite{PDG} are reproduced 
by the Schr\"odinger equation, while the up and 
down quark masses are 0.32 GeV and the strange quark mass is 0.5 GeV 
\cite{JSX}. The temperature-dependent potential between constituents $a$
and $b$ is given by \cite{JSX}
 \begin{equation}
V_{ab}(\vec {r}) = V_{\rm si}(\vec {r}) + V_{\rm ss}(\vec {r}),\label{eq12}
\end{equation}
where $\vec{r}$ is the relative coordinate of $a$ and $b$. The first term 
$V_{\rm si}(\vec{r})$ is the central spin-independent potential and depends on 
temperature:
\begin{equation}
V_{\rm {si}}(\vec {r}) =
- \frac {\vec{\lambda}_a}{2} \cdot \frac {\vec{\lambda}_b}{2}
\frac {3}{4} D \left[ 1.3- \left( \frac {T}{T_{\rm c}} \right)^4 \right]
\tanh (Ar) + \frac {\vec{\lambda}_a}{2} \cdot \frac {\vec{\lambda}_b}{2}
\frac {6\pi}{25} \frac {v(\lambda r)}{r} \exp (-Er),
\end{equation}
where $D=0. 7$ GeV, $T_{\rm c}=0.175$ GeV,
$A=1.5[0.75+0.25 (T/{T_{\rm c}})^{10}]^6$ GeV, $E=0. 6$ GeV,
$\lambda=\sqrt{25/16\pi^2 \alpha'}$ with $\alpha'=1.04$ GeV$^{-2}$,
$\vec{\lambda}_a$ ($\vec{\lambda}_b$) are the Gell-Mann matrices for the color
generators of constituent $a~(b)$, and the dimensionless function $v(x)$ is
given by Buchm\"uller and Tye in Ref. \cite{BT}.
	
This potential $V_{\rm si}(\vec{r})$ is relevant to the temperature of 
hadronic matter and the distance $r$.
It shows some characteristics as follows. At very short distances $r < 0.01$ 
fm, the potential arises from one-gluon exchange plus perturbative one- and 
two-loop corrections. At large distances and blow the 
QCD phase-transition temperature $T_{\rm c}$, the color screening produced by 
high-temperature medium may be strong. Karsch et al. \cite{KLP} have
provided such a numerical quark-antiquark potential at $r \geq 0.3$ fm
in a temperature region from lattice QCD calculations. When the distance 
between the quark and the antiquark becomes large, the quark-antiquark 
potential at a given temperature becomes a constant value, which decreases with
increasing temperature.

The second term $V_{\rm ss}(\vec{r})$ in Eq. (9) is the spin-spin interaction
which originates from one-gluon exchange plus perturbative one- and two-loop 
corrections \cite{Xu}, depends on constituent masses,
and includes relativistic effects \cite{BS,GI}:
\begin{eqnarray}
V_{\rm ss}(\vec {r})=
-\frac {\vec{\lambda}_a}{2} \cdot \frac {\vec{\lambda}_b}{2}
\frac {16\pi^2}{25}\frac{d^3}{\pi^{3/2}}\exp(-d^2r^2) \frac {\vec {s}_a \cdot
\vec {s} _b} {m_am_b}
+ \frac {\vec{\lambda}_a}{2} \cdot \frac {\vec{\lambda}_b}{2}\frac {4\pi}{25}
\frac {1} {r}
\frac {d^2v(\lambda r)}{dr^2} \frac {\vec {s}_a \cdot \vec {s}_b}{m_am_b} ,
\end{eqnarray}
where $\vec{s}_a$ ($\vec{s}_b$) is the spin of 
constituent $a$ ($b$), and $d$ is given by
\begin{eqnarray}
d^2=d_1^2\left[\frac{1}{2}+\frac{1}{2}\left(\frac{4m_a m_b}{(m_a+m_b)^2}
\right)^4\right]+d_2^2\left(\frac{2m_am_b}{m_a+m_b}\right)^2,
\end{eqnarray}
where $d_1=0.15$ GeV and $d_2= 0.705$.

\vspace{0.5cm}
\leftline{\bf IV. NUMERICAL CROSS SECTIONS AND DISCUSSIONS }
\vspace{0.5cm}

We consider the following inelastic meson-meson scattering processes that 
mainly take the two Feynman diagrams in Fig. 1:
\begin{eqnarray}
\nonumber
K \bar {K} \to K \bar {K}^\ast , \quad K \bar{K} \to K^* \bar{K},
\\  \nonumber
\pi K \to \pi K^\ast, \quad \pi K \to \rho K,
\\  \nonumber
\pi \pi \to K \bar{K}^\ast, \quad \pi \pi \to K^\ast \bar{K},
\quad \pi \pi \to K^\ast \bar{K}^\ast,
\\  \nonumber
\pi \rho \to K \bar{K}, \quad \pi \rho \to K^\ast \bar{K}^\ast,
\\  \nonumber
\rho \rho \to K^\ast \bar{K}^\ast, \quad K \bar{K}^\ast \to \rho \rho, 
\quad K^* \bar{K} \to \rho \rho.
\end{eqnarray}
The total spin of $\pi$ and $K$ mesons does not equal the total spin of $\pi$
and $K^*$ mesons or of $\rho$ and $K$ mesons, i.e., 
the total spin in either $\pi K \to \pi K^*$ or
$\pi K \to \rho K$ is not conserved. Quark interchange thus does
not happen in the two reactions. ${\cal M}_{{\rm a}q_1\bar {q}_2}$ and 
${\cal M}_{{\rm a}\bar{q}_1q_2}$ are proportional to the flavor matrix 
elements. If the transition amplitudes equal zero, the unpolarized cross
section given in Eq. (1) is zero. As seen in Table 1, the flavor matrix 
elements for the two reactions for $I=3/2$
are zero. Quark-antiquark annihilation and creation
does not happen in the two reactions for $I=3/2$ too. Therefore, cross sections
for $\pi K \to \pi K^*$ for $I=3/2$ and $\pi K \to \rho K$ for $I=3/2$ are zero
in the present work, but we still investigate $\pi K \to \pi K^*$ for $I=1/2$ 
and $\pi K \to \rho K$ for $I=1/2$ of which ${\cal M}_{{\rm a}\bar{q}_1q_2f}$
are not zero. The other reactions must involve
quark-antiquark annihilation and creation, but do not involve quark 
interchange.

It is shown in Table~\ref{table1} that only the right diagram in 
Fig.~\ref{fig1}
contributes to the reactions: $\pi \pi \to K \bar{K}^*$,
~$\pi \pi \to K^\ast \bar{K}$, ~$\pi \pi \to K^\ast \bar{K}^*$,
~$\pi \rho \to K \bar{K}$, ~$\pi \rho \to K^* \bar{K}^\ast$,
~$\rho \rho \to K^* \bar{K}^*$, ~$K \bar{K}^* \to \rho \rho$, and
~$K^\ast \bar{K} \to \rho \rho$. Since the flavor matrix elements for the
reactions for $I=0$ are $\sqrt {6}/2$ times the ones for $I=1$,
the cross sections for the reactions for $I=0$ are 1.5 times the cross sections
for $I=1$. The cross section for $K\bar{K} \to K^*\bar{K}$
($\pi \pi \to K^* \bar{K}$, $K^*\bar{K} \to \rho\rho$) equals the one for 
$K\bar{K} \to K\bar{K}^*$ ($\pi \pi \to K\bar{K}^*$,
$K\bar{K}^* \to \rho \rho$). We thus do not plot the cross sections for
$K\bar{K} \to K^*\bar{K}$,
$\pi \pi \to K^* \bar{K}$, and $K^*\bar{K} \to \rho \rho$.

The gauge coupling constant is $\frac{2\sqrt{6}\pi}{5}$ for quark-antiquark
annihilation and creation \cite{YXW,BT}. According to Eq. (1),
we calculate unpolarized cross sections at the six temperatures
$T/T_{\rm c} =0$, $0.65$, $0.75$, $0.85$, $0.9$, and $0.95$. In 
Figs.~\ref{fig3}-\ref{fig12}
we plot the unpolarized cross sections for the following ten channels:
\begin{eqnarray}
\nonumber I=1~ K \bar {K} \to K \bar {K}^\ast , \quad I=0~ K
\bar{K} \to K \bar{K}^\ast,
\\  \nonumber
I=1/2~ \pi K \to \pi K^\ast, \quad I=1/2~ \pi K \to \rho K,
\\  \nonumber
I=1~ \pi \pi \to K \bar{K}^\ast, \quad I=1~ \pi \pi \to K^\ast
\bar{K}^\ast,
\\  \nonumber
I=1~ \pi \rho \to K \bar{K}, \quad I=1~ \pi \rho \to K^\ast
\bar{K}^\ast,
\\  \nonumber
I=1~ \rho \rho \to K^\ast \bar{K}^\ast, \quad I=1~ K \bar{K}^\ast \to \rho\rho.
\end{eqnarray}
The last channel is endothermic at $T/T_{\rm c}=0$ and exothermic at
$T/T_{\rm c}=0.65$, 0.75, 0.85, 0.9, and 0.95. The other nine channels are
endothermic.
The numerical cross sections for endothermic reactions are parametrized as
\begin{eqnarray}
\sigma^{\rm unpol}(\sqrt {s},T)
&=&a_1 \left( \frac {\sqrt {s} -\sqrt {s_0}} {b_1} \right)^{e_1}
\exp \left[ e_1 \left( 1-\frac {\sqrt {s} -\sqrt {s_0}} {b_1} \right) \right]
\nonumber \\
&&+ a_2 \left( \frac {\sqrt {s} -\sqrt {s_0}} {b_2} \right)^{e_2}
\exp \left[ e_2 \left( 1-\frac {\sqrt {s} -\sqrt {s_0}} {b_2} \right) \right],
\end{eqnarray}
where $\sqrt{s_0}$ is the threshold energy, and $a_1$, $b_1$, $e_1$, $a_2$,
$b_2$, and $e_2$ are parameters. The numerical cross sections
for exothermic reactions are parametrized as
\begin{eqnarray}
\sigma^{\rm unpol}(\sqrt {s},T)
&=&\frac{\vec{P}^{\prime 2}}{\vec{P}^2}
\left\{a_1 \left( \frac {\sqrt {s} -\sqrt {s_0}} {b_1} \right)^{e_1}
\exp \left[ e_1 \left( 1-\frac {\sqrt {s} -\sqrt {s_0}} {b_1} \right) \right]
\right.
\nonumber \\
&&+ \left.
a_2 \left( \frac {\sqrt {s} -\sqrt {s_0}} {b_2} \right)^{e_2}
\exp \left[ e_2 \left( 1-\frac {\sqrt {s} -\sqrt {s_0}} {b_2} \right) \right]
\right\}.
\end{eqnarray}
The parameter values are listed in Tables 2-4. In the three tables
the quantity $d_0$ is the separation 
between the peak's location on the $\sqrt s$-axis and the threshold energy. 
The smaller $d_0$ is, the faster the cross section increases from zero to the
peak cross section. The quantity $\sqrt{s_z}$ is the square root of the 
Mandelstam variable at which the cross section is 1/100 of the peak cross 
section. The quantity
$\sqrt{s_z}-\sqrt{s_0}-d_0$ is the difference between $\sqrt{s_z}$ and the
peak's location on the $\sqrt{s}$-axis. The smaller 
$\sqrt{s_z}-\sqrt{s_0}-d_0$ is, the faster the cross section decreases from
the peak cross section to zero.

The potential given in Eq. (9) depends on temperature. The Schr\"odinger
equation with the potential yields temperature-dependent meson masses. For
any endothermic (exothermic) 2-to-2 meson-meson reaction the threshold energy
is the sum of the masses of the two final (initial) mesons. Since the meson 
masses decrease with increasing temperature, the threshold energy decreases
with increasing temperature.

Let $m_A$ ($m_B$, $m_C$, $m_D$) be the mass of meson $A$ ($B$, $C$, $D$). In 
terms of the meson masses we have
\begin{displaymath}
\mid \vec{P} \mid =\frac {1}{2} 
\sqrt {\frac {(s-m_A^2-m_B^2)^2-4m_A^2m_B^2}{s}},
\end{displaymath}
\begin{displaymath}
\mid \vec{P}^\prime \mid =\frac {1}{2} 
\sqrt {\frac {(s-m_C^2-m_D^2)^2-4m_C^2m_D^2}{s}}.
\end{displaymath}
For endothermic reactions the threshold energy is $\sqrt{s_0}=m_C+m_D$, and 
we obtain
\begin{displaymath}
\frac {\mid \vec{P}^\prime \mid}{\mid \vec{P} \mid} =
\sqrt {\frac {(\sqrt{s}-\sqrt{s_0})(\sqrt{s}+\sqrt{s_0})[s-(m_C-m_D)^2]}
{[s-(m_A+m_B)^2][s-(m_A-m_B)^2]}}.
\end{displaymath}
When $\sqrt s$ is close to $\sqrt {s_0}$, $\frac {\mid \vec{P}^\prime \mid}
{\mid \vec{P} \mid}$ is sensitive to $\sqrt {s} - \sqrt {s_0}$. Since Eq. (1)
contains $\frac {\mid \vec{P}^\prime \mid}{\mid \vec{P} \mid}$, the
unpolarized cross section in Eq. (13) may be proportional to
$(\sqrt {s} - \sqrt {s_0})^{0.5}$. Nevertheless, the factor
$\sqrt {\frac {(\sqrt{s}+\sqrt{s_0})[s-(m_C-m_D)^2]}
{[s-(m_A+m_B)^2][s-(m_A-m_B)^2]}}$ in the expression of 
$\frac {\mid \vec{P}^\prime \mid} {\mid \vec{P} \mid}$ and the factor
$\frac {1}{s}
\mid {\cal M}_{{\rm a}q_1\bar{q}_2}+{\cal M}_{{\rm a}\bar{q}_1q_2} \mid^2$
in Eq. (1) modify the dependence of the unpolarized cross section on 
$\sqrt {s} - \sqrt {s_0}$. We thus use $(\sqrt {s} - \sqrt {s_0})^{e_1}$ and
$(\sqrt {s} - \sqrt {s_0})^{e_2}$ in Eq. (13)
instead of $(\sqrt {s} - \sqrt {s_0})^{0.5}$. Indeed, it is shown in Tables 
2-4 that the values of $e_1$ and/or $e_2$ are near 0.5.

Denote by $\sigma_{\rm num}^{\rm unpol}(\sqrt{s},T)$ the cross sections 
calculated from
Eq. (1), which are plotted in Figs. 3-12. By comparison we denote by 
$\sigma_{\rm para}^{\rm unpol}(\sqrt{s},T)$ the cross sections given by Eqs.
(13) and (14). We change $a_1$, $b_1$, $e_1$, $a_2$, $b_2$, and $e_2$ to make
$\mid (\sigma_{\rm para}^{\rm unpol}(\sqrt{s},T)
-\sigma_{\rm num}^{\rm unpol}(\sqrt{s},T))/
\sigma_{\rm num}^{\rm unpol}(\sqrt{s},T) \mid$ as small as possible. The 
parameter values which make $\mid (\sigma_{\rm para}^{\rm unpol}(\sqrt{s},T)
-\sigma_{\rm num}^{\rm unpol}(\sqrt{s},T))/
\sigma_{\rm num}^{\rm unpol}(\sqrt{s},T) \mid$ smallest are provided in
Tables 2-4.

A feature of the endothermic reactions in Figs. 3-11 is that 
the peak cross section decreases first and then increases as the
temperature goes up. As the temperature increases from zero, 
confinement shown by the potential in Eq. (10) becomes weaker and weaker,
the Schr\"odinger equation produces increasing meson radii, and mesonic
quark-antiquark states become looser and looser. On one hand the weakening
confinement with increasing temperature makes combining final 
quarks and antiquarks into final mesons more difficult, and thus reduces
cross sections; On the other hand the increasing radii of initial
mesons cause increasing cross sections as the temperature goes up. 
The two factors determine the change in peak cross section with respect to the
temperature. Another feature is that the cross section increases rapidly from
zero to a maximum value when the total energy of the two initial mesons in the
center-of-mass frame increases from the threshold energy, and the cross
section further decreases from the maximum value or exhibits a plateau
on the right of the peak as seen in Figs. 8 and 10.

The unpolarized cross sections for $K \bar{K} \to K \bar{K}^\ast$ for $I=1$ and
for $I=0$ are shown in Figs.~\ref{fig3} and \ref{fig4}, respectively. 
Quark-antiquark annihilation and creation takes place in the two 
isospin channels.
If ${\cal M}_{{\rm a}q_1\bar {q}_2}$ (${\cal M}_{{\rm a}\bar{q}_1q_2}$) for
a reaction equals zero, the left (right) diagram does not contribute to the
reaction.
It is shown from Table~\ref{table1} that the two diagrams in Fig.~\ref{fig1} 
contribute to $K \bar{K} \to K \bar{K}^\ast$ for $I=0$, and only the right 
diagram contributes to $K \bar{K} \to K \bar{K}^\ast$ for $I=1$.
The peak cross section of 
$K \bar{K} \to K \bar{K}^\ast$ for $I=0$ at a given temperature is more than 4 
times the one for $I=1$, and $\sqrt{s_z}$ for $I=0$ in Table 2 is roughly 2
times that for 
$I=1$. When $T/T_{\rm c} =0$, $0.65$, and $0.75$, $d_0$ of $K \bar{K} \to K 
\bar{K}^\ast$ for $I=0$ is equal to $d_0$ for $I=1$; when $T/T_{\rm c}=0.85$,
$0.90$, and $0.95$, $d_0$ of $K \bar{K} \to K \bar{K}^\ast$ for $I=0$ is less 
than $d_0$ for $I=1$. Therefore, the cross section for 
$K\bar{K} \to K\bar{K}^*$ for $I=0$ is larger than the one for 
$K\bar{K} \to K\bar{K}^*$ for $I=1$.

The unpolarized cross section for $K\bar{K}\to\rho\rho$ for $I=1$ has been 
shown in Fig.~15 of Ref.\cite{SXW}, and the one for $K\bar{K}^{\ast}\to 
\rho\rho$ for $I=1$ in Fig. 12 in the present work. At zero temperature 
the peak cross
section of $K\bar{K}\to\rho\rho$ for $I=1$ is smaller than the one of
$K\bar{K}^*\to\rho\rho$ for $I=1$; the cross section for 
$K\bar{K}\to\rho\rho$ for $I=1$ decreases from the peak cross section
slower than for
$K\bar{K}^*\to\rho\rho$ for $I=1$ since $\sqrt s_z$ of the former is larger
than that of the latter. When the two reactions are exothermic, the cross 
section for $K\bar{K}\to\rho\rho$ for $I=1$ 
decreases slower than for $K\bar{K}^*\to\rho\rho$ for $I=1$ with
increasing center-of-mass energy of the two initial mesons
from the threshold energy plus $10^{-4}$ GeV. 

In the present work and in Ref. \cite{SXW} we have studied the reactions:
$K\bar{K} \to K\bar{K}^*$, $K^*\bar{K}$, and $K^*\bar{K}^*$; 
$\pi\pi \to K\bar{K}$,
$K\bar{K}^*$, $K^*\bar{K}$, and $K^*\bar{K}^*$; $\pi\rho \to K\bar{K}$,
$K\bar{K}^*$, $K^*\bar{K}$, and $K^*\bar{K}^*$. As an example we compare the
production of $K\bar{K}^*$ with the production of $K^*\bar{K}^*$ in the
$K+\bar K$ reaction. The unpolarized cross sections for $K\bar{K} \to K^{\ast} 
\bar{K}^\ast$ for $I=1$ and for $I=0$ have been shown in Figs. 8 and 9 of 
Ref.~\cite{SXW}, respectively. Quark-antiquark annihilation and 
creation takes place in the two isospin channels. 
The peak cross section of $K \bar{K} \to K \bar{K}^\ast$ for 
$I=1$ ( $I=0$ ) at a given temperature is larger than the one of 
$K \bar{K} \to K^* \bar{K}^\ast$ for $I=1$ ( $I=0$ ).
The Mandelstam variable $\sqrt s$ corresponding to the peak cross section of
$K\bar{K} \to K\bar{K}^*$ is smaller than the one corresponding to the peak 
cross section of $K\bar{K} \to K^*\bar{K}^*$ at a given temperature. Since the
two reactions have the same initial mesons, the initial mesons in 
$K\bar{K} \to K\bar{K}^*$ have a smaller value of $\mid{\vec{P}}\mid$ 
corresponding to the peak cross section than
in $K \bar{K} \to K^{\ast} \bar{K}^\ast$. The cross section given in Eq. (1)
is proportional to the inverse of $s\mid{\vec{P}}\mid$.
Therefore, the peak cross section of 
$K \bar{K} \to K \bar{K}^\ast$ is larger than the one
of $K \bar{K} \to K^{\ast} \bar{K}^\ast$. 

The total spin of the two final mesons may not equal the total spin of the two
initial mesons in the reactions in the present work. This can be accounted for
from a quark-antiquark potential which is equivalent to the transtion potential
in Eq. (4). The quark-antiquark potential is obtained from 
the transition potential under the Fierz transformations in Ref. \cite{SXW},
\begin{eqnarray}
V_{{\rm a}q\bar{q}\rm F}(\vec{k})&=&-\frac{g^{2}_s}{k^2}
\left[\frac{1}{3}\lambda^{0}_{f}(31)\lambda^{0}_{f}(42)
+\frac{1}{2}\vec{\lambda}_{f}(31)\cdot\vec{\lambda}^{T}_{f}(42)\right]
 \nonumber     \\
&\times&\left[\frac{4}{9}\lambda^{0}(31)\lambda^{0}(42)
-\frac{1}{12}\vec{\lambda}(31)\cdot\vec{\lambda}^{T}(42)\right]
\left[-\frac{3}{2}-\frac{1}{2}\vec{\sigma}(31)\cdot\vec{\sigma}(42) \right.
\nonumber     \\
&+&\frac{\vec{\sigma}(42)\cdot\vec{p}_{4} \vec{\sigma}(42)\cdot\vec{p}_{2}}
{8m_{q}m_{q'}}
+\frac{\vec{\sigma}(31)\cdot\vec{p}_{3} \vec{\sigma}(31)\cdot\vec{p}_{1}}
{8m_{q}m_{q'}}
+\frac{3\vec{\sigma}(31)\cdot\vec{p}_{1} \vec{\sigma}(42)\cdot\vec{p}_{2}}
{8m^{2}_{q}}
\nonumber     \\
&-&\frac{\vec{\sigma}(31)\cdot\vec{p}_{1} \vec{\sigma}(42)\cdot\vec{p}_{4}}
{8m_{q}m_{q'}}
-\frac{\vec{\sigma}(31)\cdot\vec{p}_{3} \vec{\sigma}(42)\cdot\vec{p}_{2}}
{8m_{q}m_{q'}}
+\frac{3\vec{\sigma}(31)\cdot\vec{p}_{3} \vec{\sigma}(42)\cdot\vec{p}_{4}}
{8m^{2}_{q'}}
\nonumber     \\
&+&\frac{\vec{\sigma}(31)\vec{\sigma}(31)\cdot\vec{p}_{1} 
\vec{\sigma}(42)\vec{\sigma}(42)\cdot\vec{p}_{2}}{8m^{2}_{q}}
+\frac{\vec{\sigma}(31)\vec{\sigma}(31)\cdot\vec{p}_{1} 
\vec{\sigma}(42)\cdot\vec{p}_{4}\vec{\sigma}(42)}{8m_{q}m_{q'}}
 \nonumber     \\
&+&\frac{\vec{\sigma}(31)\cdot\vec{p}_{3}\vec{\sigma}(31)\cdot 
\vec{\sigma}(42)\vec{\sigma}(42)\cdot\vec{p}_{2}}{8m_{q}m_{q'}} 
+\frac{\vec{\sigma}(31)\cdot\vec{p}_{3}\vec{\sigma}(31)
\vec{\sigma}(42)\cdot\vec{p}_{4}\vec{\sigma}(42)}{8m^{2}_{q'}}
 \nonumber     \\
&-&\left. \frac{\vec{\sigma}(42)\cdot\vec{p}_{4}\vec{\sigma}(31)\cdot 
\vec{\sigma}(42)\vec{\sigma}(42)\cdot\vec{p}_{2}}{8m_{q}m_{q'}}
-\frac{\vec{\sigma}(31)\cdot\vec{p}_{3}\vec{\sigma}(31)\cdot 
\vec{\sigma}(42)\vec{\sigma}(31)\cdot\vec{p}_{1}}{8m_{q}m_{q'}} \right],
 \nonumber     \\
\end{eqnarray}
where $\lambda^0_f$ is a $3\times3$ unit matrix in flavor space, 
$\vec{\lambda}_f$
are the Gell-Mann matrices that operate in flavor space, and $\lambda^0$ is 
a $3\times3$ unit matrix in color space. The superscript $T$ in Eqs. (15),
(16), and (19)-(21) means transposition \cite{AW}.

For the left diagram in Fig.~\ref{fig1}, the quark-antiquark potential 
corresponding to $q_1+\bar q_2 \to q_3 + \bar q_4$ is between $q_1$ and 
$\bar{q}_2$,
\begin{eqnarray}
V_{{\rm a}q_1\bar{q}_2\rm F}(\vec{k})
&=&-\frac{g^{2}_s}{k^2}\left[\frac{1}{3}\lambda^{0}_{f}(31)\lambda^{0}_{f}(42)
+\frac{1}{2}\vec{\lambda}_{f}(31)\cdot\vec{\lambda}^{T}_{f}(42)\right]
 \nonumber     \\
&\times&\left[\frac{4}{9}\lambda^{0}(31)\lambda^{0}(42)-\frac{1}{12}
\vec{\lambda}(31)\cdot\vec{\lambda}^{T}(42)\right]
\left[-\frac{3}{2}-\frac{1}{2}
\vec{\sigma}(31)\cdot\vec{\sigma}(42) \right.
\nonumber     \\
&+&\frac{\vec{\sigma}(42)\cdot\vec{p}_{{\bar q_4}}\vec{\sigma}(42)\cdot
\vec{p}_{{{\bar q_2}}}}{8m_{q_1}m_{q_3}}
+\frac{\vec{\sigma}(31)\cdot\vec{p}_{q_3}\vec{\sigma}(31)\cdot
\vec{p}_{q_1}}{8m_{q_1}m_{q_3}}
+\frac{3\vec{\sigma}(31)\cdot\vec{p}_{q_1} \vec{\sigma}(42)\cdot
\vec{p}_{\bar q_2}}{8m^{2}_{q_1}}
\nonumber     \\
&-&\frac{\vec{\sigma}(31)\cdot\vec{p}_{q_1} \vec{\sigma}(42)\cdot
\vec{p}_{\bar q_4}}{8m_{q_1}m_{q_3}}
-\frac{\vec{\sigma}(31)\cdot\vec{p}_{q_3} \vec{\sigma}(42)\cdot
\vec{p}_{\bar q_2}}{8m_{q_1}m_{q_3}}
+\frac{3\vec{\sigma}(31)\cdot\vec{p}_{q_3} \vec{\sigma}(42)\cdot
\vec{p}_{\bar q_4}}{8m^{2}_{q_3}}
\nonumber     \\
&+&\frac{\vec{\sigma}(31)\vec{\sigma}(31)\cdot\vec{p}_{q_1} 
\vec{\sigma}(42)\vec{\sigma}(42)\cdot\vec{p}_{\bar q_2}}{8m^{2}_{q_1}}
+\frac{\vec{\sigma}(31)\vec{\sigma}(31)\cdot\vec{p}_{q_1} 
\vec{\sigma}(42)\cdot\vec{p}_{\bar q_4}\vec{\sigma}(42)}{8m_{q_1}m_{q_3}}
 \nonumber     \\
&+&\frac{\vec{\sigma}(31)\cdot\vec{p}_{q_3}\vec{\sigma}(31)\cdot 
\vec{\sigma}(42)\vec{\sigma}(42)\cdot\vec{p}_{\bar q_2}}{8m_{q_1}m_{q_3}} 
+\frac{\vec{\sigma}(31)\cdot\vec{p}_{q_3}\vec{\sigma}(31)\vec{\sigma}(42)
\cdot\vec{p}_{\bar q_4}\vec{\sigma}(42)}{8m^{2}_{q_3}}
 \nonumber     \\
&-&\left. \frac{\vec{\sigma}(42)\cdot\vec{p}_{\bar q_4}\vec{\sigma}(31)\cdot 
\vec{\sigma}(42)\vec{\sigma}(42)\cdot\vec{p}_{\bar q_2}}{8m_{q_1}m_{q_3}}
-\frac{\vec{\sigma}(31)\cdot\vec{p}_{q_3}\vec{\sigma}(31)\cdot 
\vec{\sigma}(42)\vec{\sigma}(31)\cdot\vec{p}_{q_1}}{8m_{q_1}m_{q_3}}\right] .
 \nonumber     \\
\end{eqnarray}
While we carry out the Fierz transformations, we need that quark 3 (antiquark
4) and quark 1 (antiquark 2) have the same flavor. Then, quark and antiquark
masses possess
$m_{q_3}=m_{q_1}$ and $m_{\bar{q}_4}=m_{\bar{q}_2}$. For convenience
we use $\vec{p}_{q_1}^{~\prime}=\vec{p}_{q_3}$ and 
$\vec{p}_{\bar{q}_2}^{~\prime}=\vec{p}_{\bar{q}_4}$.
The Hamiltonian corresponding to the left diagram in Fig.~\ref{fig1} is
\begin{eqnarray}
H_1 &=& V_{{\rm a}q_{1}\bar{q}_{2}\rm F}(\vec{k})+V_{q_1\bar{q}_1}
+V_{q_2\bar{q}_2}
\nonumber     \\
&+&\sqrt {m^2_{q_1}+\vec{p}^{~2}_{q_1}}
+\sqrt {m^2_{\bar{q}_1}+\vec{p}^{~2}_{\bar{q}_1}}
+\sqrt {m^2_{q_2}+\vec{p}^{~2}_{q_2}}
+\sqrt {m^2_{\bar{q}_2}+\vec{p}^{~2}_{\bar{q}_2}} ,
\end{eqnarray}
where $V_{q_1\bar{q}_1}$ and $V_{q_2\bar{q}_2}$ are the Fourier transform
of $V_{ab}(\vec{r})$ in Eq. (9). The total spin of the two mesons is
\begin{eqnarray}
\vec{S}=\vec{S}_{q_1}+\vec{S}_{\bar{q}_1}+\vec{S}_{q_2}+\vec{S}_{\bar{q}_2} .
\end{eqnarray}

The commutator of the $z$ component $S_z$ of the total spin and the
Hamiltonian is
\begin{eqnarray}
[S_z,H_1]&=&-\frac{g^{2}_s}{k^2} \frac{1}{8m^{2}_{q_1}}
\left(\frac{1}{3}\lambda^{0}_{q_1f}\lambda^{0}_{\bar{q}_2f}
+\frac{1}{2}\vec{\lambda}_{q_1f}\cdot\vec{\lambda}^{T}_{\bar {q}_2f}\right)
\left(\frac{4}{9}\lambda^{0}_{q_1}\lambda^{0}_{\bar{q}_2}
-\frac{1}{12}\vec{\lambda}_{q_1}\cdot\vec{\lambda}^{T}_{\bar {q}_2}\right)
\nonumber     \\
    &\times& \{\sigma_{\bar {q}_{2}y}[p'_{\bar {q}_{2}z}p_{\bar
{q}_{2}y}-p'_{\bar {q}_{2}y}p_{\bar {q}_{2}z} -p_{q_{1}y}p_{\bar
{q}_{2}z}+p_{q_{1}z}p_{\bar {q}_{2}y}+p_{q_{1}y}p'_{\bar {q}_{2}z}
-p_{q_{1}z}p'_{\bar {q}_{2}y}-p'_{q_{1}y}p_{\bar {q}_{2}z}
\nonumber     \\
    &+&p'_{q_{1}z}p_{\bar {q}_{2}y}+p'_{q_{1}y}p'_{\bar {q}_{2}z}
-p'_{q_{1}z}p'_{\bar{q}_{2}y}+p'_{q_{1}z}p_{q_{1}y}-p'_{q_{1}y}p_{q_{1}z}]
\nonumber     \\
    &+&\sigma_{\bar {q}_{2}x}[p'_{\bar {q}_{2}z}p_{\bar
{q}_{2}x}-p'_{\bar {q}_{2}x}p_{\bar {q}_{2}z} -p_{q_{1}x}p_{\bar
{q}_{2}z}+p_{q_{1}z}p_{\bar {q}_{2}x}+p_{q_{1}x}p'_{\bar {q}_{2}z}
-p_{q_{1}z}p'_{\bar {q}_{2}x}-p'_{q_{1}x}p_{\bar {q}_{2}z}
\nonumber     \\
    &+&p'_{q_{1}z}p_{\bar {q}_{2}x}+p'_{q_{1}x}p'_{\bar {q}_{2}z}
-p'_{q_{1}z}p'_{\bar{q}_{2}x}+p'_{q_{1}z}p_{q_{1}x}-p'_{q_{1}x}p_{q_{1}z}]
\nonumber     \\
    &+&\sigma_{q_{1}y}[p'_{q_{1}z}p_{q_{1}y}-p'_{q_{1}y}p_{q_{1}z}
-p_{q_{1}z}p_{\bar {q}_{2}y}+p_{q_{1}y}p_{\bar
{q}_{2}z}-p_{q_{1}z}p'_{\bar {q}_{2}y} +p_{q_{1}y}p'_{\bar
{q}_{2}z}+p'_{q_{1}z}p_{\bar {q}_{2}y}
\nonumber     \\
    &-&p'_{q_{1}y}p_{\bar {q}_{2}z}+p'_{q_{1}z}p'_{\bar {q}_{2}y}
-p'_{q_{1}y}p'_{\bar {q}_{2}z}+p'_{\bar {q}_{2}z}p_{\bar
{q}_{2}y}-p'_{\bar {q}_{2}y}p_{\bar {q}_{2}z}]
\nonumber     \\
    &+&\sigma_{q_{1}x}[p'_{q_{1}z}p_{q_{1}x}-p'_{q_{1}x}p_{q_{1}z}
-p_{q_{1}z}p_{\bar {q}_{2}x}+p_{q_{1}x}p_{\bar
{q}_{2}z}-p_{q_{1}z}p'_{\bar {q}_{2}x} +p_{q_{1}x}p'_{\bar
{q}_{2}z}+p'_{q_{1}z}p_{\bar {q}_{2}x}
\nonumber     \\
    &-&p'_{q_{1}x}p_{\bar {q}_{2}z}+p'_{q_{1}z}p'_{\bar {q}_{2}x}
-p'_{q_{1}x}p'_{\bar {q}_{2}z} +p'_{\bar
{q}_{2}z}p_{\bar {q}_{2}x}-p'_{\bar {q}_{2}x}p_{\bar {q}_{2}z}]
\nonumber     \\
    &+&i\sigma_{q_{1}x}\sigma_{\bar {q}_{2}z}[-3p_{q_{1}y}p_{\bar
{q}_{2}z} +p_{q_{1}y}p'_{\bar {q}_{2}z}+p'_{q_{1}y}p_{\bar
{q}_{2}z}-3p'_{q_{1}y}p'_{\bar {q}_{2}z} -p_{q_{1}z}p_{\bar
{q}_{2}y}+p'_{q_{1}z}p_{\bar {q}_{2}y}
\nonumber     \\
    &+&p_{q_{1}z}p'_{\bar{q}_{2}y}-p'_{q_{1}z}p'_{\bar {q}_{2}y}
+p'_{\bar {q}_{2}y}p_{\bar {q}_{2}z}
+p'_{\bar {q}_{2}z}p_{\bar {q}_{2}y}+p'_{q_{1}z}p_{q_{1}y} 
+p'_{q_{1}y}p_{q_{1}z}]
\nonumber     \\
&+&i\sigma_{q_{1}y}\sigma_{\bar {q}_{2}z}[3p_{q_{1}x}p_{\bar
{q}_{2}z} -p_{q_{1}x}p'_{\bar {q}_{2}z}-p'_{q_{1}x}p_{\bar
{q}_{2}z}+3p'_{q_{1}x}p'_{\bar {q}_{2}z} +p_{q_{1}z}p_{\bar
{q}_{2}x}-p_{q_{1}z}p'_{\bar {q}_{2}x}
\nonumber     \\
    &-&p'_{q_{1}z}p_{\bar{q}_{2}x}+p'_{q_{1}z}p'_{\bar {q}_{2}x}
-p'_{\bar {q}_{2}z}p_{\bar {q}_{2}x}-p'_{\bar {q}_{2}x}p_{\bar
{q}_{2}z}-p'_{q_{1}z}p_{q_{1}x} -p'_{q_{1}x}p_{q_{1}z}]
\nonumber     \\
    &+&i(\sigma_{q_{1}x}\sigma_{\bar {q}_{2}y}
+\sigma_{q_{1}y}\sigma_{\bar {q}_{2}x}) [4p_{q_{1}x}p_{\bar
{q}_{2}x}-4p_{q_{1}y}p_{\bar {q}_{2}y} -2p_{q_{1}x}p'_{\bar
{q}_{2}x}+2p_{q_{1}y}p'_{\bar {q}_{2}y}
\nonumber     \\
    &-&2p'_{q_{1}x}p_{\bar{q}_{2}x} +2p'_{q_{1}y}p_{\bar {q}_{2}y}
+4p'_{q_{1}x}p'_{\bar {q}_{2}x}-4p'_{q_{1}y}p'_{\bar {q}_{2}y}
-2p'_{\bar {q}_{2}x}p_{\bar {q}_{2}x} +2p'_{\bar {q}_{2}y}p_{\bar {q}_{2}y}
\nonumber     \\
    &-&2p'_{q_{1}x}p_{q_{1}x}+2p'_{q_{1}y}p_{q_{1}y}]
\nonumber     \\
    &+&i(\sigma_{q_{1}y}\sigma_{\bar {q}_{2}y}
-\sigma_{q_{1}x}\sigma_{\bar {q}_{2}x})
[4p_{q_{1}y}p_{\bar{q}_{2}x}+4p_{q_{1}x}p_{\bar {q}_{2}y} -2p_{q_{1}y}p'_{\bar
{q}_{2}x}-2p_{q_{1}x}p'_{\bar {q}_{2}y}
\nonumber     \\
    &-&2p'_{q_{1}y}p_{\bar{q}_{2}x} -2p'_{q_{1}x}p_{\bar {q}_{2}y}
+4p'_{q_{1}y}p'_{\bar {q}_{2}x}+4p'_{q_{1}x}p'_{\bar {q}_{2}y}
-2p'_{\bar {q}_{2}y}p_{\bar {q}_{2}x}-2p'_{\bar {q}_{2}x}p_{\bar{q}_2y}
\nonumber     \\
    &-&2p'_{q_{1}y}p_{q_{1}x}-2p'_{q_{1}x}p_{q_{1}y}]
\nonumber     \\
    &+&i\sigma_{q_{1}z}\sigma_{\bar {q}_{2}y}[3p_{q_{1}z}p_{\bar
{q}_{2}x}-p_{q_{1}z}p'_{\bar {q}_{2}x} -p'_{q_{1}z}p_{\bar
{q}_{2}x}+3p'_{q_{1}z}p'_{\bar {q}_{2}x}+p_{q_{1}x}p_{\bar
{q}_{2}z} -p_{q_{1}x}p'_{\bar {q}_{2}z}
\nonumber     \\
    &-&p'_{q_{1}x}p_{\bar{q}_{2}z}+p'_{q_{1}x}p'_{\bar {q}_{2}z}
-p'_{\bar {q}_{2}z}p_{\bar {q}_{2}x}-p'_{\bar {q}_{2}x}p_{\bar {q}_{2}z}
-p'_{q_{1}z}p_{q_{1}x}-p'_{q_{1}x}p_{q_{1}z}]
\nonumber     \\
    &+&i\sigma_{q_{1}z}\sigma_{\bar {q}_{2}x}[-3p_{q_{1}z}p_{\bar
{q}_{2}y}+p_{q_{1}z}p'_{\bar {q}_{2}y} +p'_{q_{1}z}p_{\bar
{q}_{2}y}-3p'_{q_{1}z}p'_{\bar {q}_{2}y}-p_{q_{1}y}p_{\bar
{q}_{2}z} +p_{q_{1}y}p'_{\bar {q}_{2}z}
\nonumber     \\
    &+&p'_{q_{1}y}p_{\bar{q}_{2}z}-p'_{q_{1}y}p'_{\bar {q}_{2}z}
+p'_{\bar {q}_{2}z}p_{\bar {q}_{2}y}+p'_{\bar {q}_{2}y}p_{\bar {q}_{2}z}
+p'_{q_{1}z}p_{q_{1}y}+p'_{q_{1}y}p_{q_{1}z}]\} ,
\nonumber     \\
\end{eqnarray}
where $p_{ix}$, $p_{iy}$, and $ p_{iz}$ are the three components of the 
momentum of incoming quark $i$ ($i=q_1, \bar{q}_2$); $p'_{ix}$, $p'_{iy}$, and
$p'_{iz}$ are the three components of the momentum of outgoing quark $i$ 
($ i=q_1, \bar{q}_2 $); $\sigma_{q_1x}$, $\sigma_{q_1y}$, and $\sigma_{q_1z}$ 
are the three components of $\vec{\sigma}_{q_1} \equiv \vec{\sigma}(31)$; 
$\sigma_{\bar{q}_2x}$, $\sigma_{\bar{q}_2y}$, and $\sigma_{\bar{q}_2z}$ 
are the three components of $\vec{\sigma}_{\bar{q}_2} \equiv \vec{\sigma}(42)$;
$\lambda^0_{q_1f} \equiv \lambda^0_f(31)$, 
$\lambda^0_{\bar{q}_2f} \equiv \lambda^0_f(42)$,
$\vec{\lambda}_{q_1f} \equiv \vec{\lambda}_f(31)$, 
$\vec{\lambda}_{\bar{q}_2f} \equiv \vec{\lambda}_f(42)$,
$\lambda^0_{q_1} \equiv \lambda^0(31)$, 
$\lambda^0_{\bar{q}_2} \equiv \lambda^0(42)$,
$\vec{\lambda}_{q_1} \equiv \vec{\lambda}(31)$, and
$\vec{\lambda}_{\bar{q}_2} \equiv \vec{\lambda}(42)$.
Applying the momentum conservation $\vec{p}_{\bar{q}_2}^{~\prime}=\vec{p}_{q_1}
+\vec{p}_{\bar{q}_2}-\vec{p}_{q_1}^{~\prime}$, we get
\begin{eqnarray}
[S_z,H_1]&=&-\frac{g^{2}_s}{k^2} \frac{1}{8m^{2}_{q_1}}
\left(\frac{1}{3}\lambda^{0}_{q_1f}\lambda^{0}_{\bar{q}_2f}
+\frac{1}{2}\vec{\lambda}_{q_1f}\cdot\vec{\lambda}^{T}_{\bar {q}_2f}\right)
\left(\frac{4}{9}\lambda^{0}_{q_1}\lambda^{0}_{\bar{q}_2}
-\frac{1}{12}\vec{\lambda}_{q_1}\cdot\vec{\lambda}^{T}_{\bar {q}_2}\right)
\nonumber     \\
&\times&\{\sigma_{\bar
{q}_{2}y}[p'_{q_{1}y}p_{q_{1}z}+p'_{q_{1}y}p_{\bar {q}_{2}z}
-p'_{q_{1}z}p_{q_{1}y}-p'_{q_{1}z}p_{\bar {q}_{2}y}+p_{\bar
{q}_{2}y}p_{q_{1}z} -p_{\bar {q}_{2}z}p_{q_{1}y}]
\nonumber     \\
&+&\sigma_{\bar {q}_{2x}}[p'_{q_{1}x}p_{q_{1}z}+p'_{q_{1}x}p_{\bar
{q}_{2}z} -p'_{q_{1}z}p_{q_{1}x}-p'_{q_{1}z}p_{\bar
{q}_{2}x}+p_{\bar {q}_{2}x}p_{q_{1}z} -p_{\bar
{q}_{2}z}p_{q_{1}x}]
\nonumber     \\
&+&\sigma_{q_{1}y}[p'_{q_{1}z}p_{q_{1}y}-p'_{q_{1}y}p_{q_{1}z}
-p_{q_{1}z}p_{\bar {q}_{2}y}+p'_{q_{1}z}p_{\bar
{q}_{2}y}+p_{q_{1}y}p_{\bar {q}_{2}z} -p'_{q_{1}y}p_{\bar
{q}_{2}z}]
\nonumber     \\
&+&\sigma_{q_{1}x}[p'_{q_{1}z}p_{q_{1}x}-p'_{q_{1}x}p_{q_{1}z}
-p_{q_{1}z}p_{\bar {q}_{2}x}+p'_{q_{1}z}p_{\bar
{q}_{2}x}+p_{q_{1}x}p_{\bar {q}_{2}z} -p'_{q_{1}x}p_{\bar
{q}_{2}z}]
\nonumber     \\
&+&i\sigma_{q_{1}x}\sigma_{\bar {q}_{2}z}[-p_{q_{1}y}p_{\bar
{q}_{2}z}-3p'_{q_{1}y}p_{q_{1}z} -3p'_{q_{1}y}p_{\bar
{q}_{2}z}+4p'_{q_{1}y}p'_{q_{1}z}-p'_{q_{1}z}p_{\bar {q}_{2}y}
\nonumber     \\
&+&p_{q_{1}z}p_{\bar {q}_{2}y}+2p_{\bar {q}_{2}z}p_{\bar {q}_{2}y}
+2p_{q_{1}y}p_{\bar {q}_{2}z}-p'_{q_{1}z}p_{q_{1}y}]
+i\sigma_{q_{1}y}\sigma_{\bar {q}_{2}z}[p_{q_{1}x}p_{\bar
{q}_{2}z}
\nonumber     \\
&+&3p'_{q_{1}x}p_{q_{1}z}+3p'_{q_{1}x}p_{\bar
{q}_{2}z}-4p'_{q_{1}x}p'_{q_{1}z} +p'_{q_{1}z}p_{\bar
{q}_{2}x}-p_{q_{1}z}p_{\bar {q}_{2}x}-2p_{\bar {q}_{2}z}p_{\bar
{q}_{2}x}
\nonumber     \\
&-&2p_{q_{1}z}p_{q_{1}x}+p'_{\bar {q}_{2}z}p_{q_{1}x}]
+i(\sigma_{q_{1}x}\sigma_{\bar {q}_{2}y}
+\sigma_{q_{1}y}\sigma_{\bar {q}_{2}x})[4p'_{q_{1}x}p_{q_{1}x}
+4p'_{q_{1}x}p_{\bar {q}_{2}x}
\nonumber     \\
&-&4p'_{q_{1}x}p'_{q_{1}x}-4p'_{q_{1}y}p_{q_{1}y}
-4p'_{q_{1}y}p_{\bar {q}_{2}y}+4p'_{q_{1}y}p'_{q_{1}y}
+2p_{q_{1}y}p_{q_{1}y}-2p_{q_{1}x}p_{q_{1}x}
\nonumber     \\
&+&2p_{\bar{q}_{2}y}p_{\bar {q}_{2}y}-2p_{\bar {q}_{2}x}p_{\bar
{q}_{2}x}]+i(\sigma_{q_{1}y}\sigma_{\bar {q}_{2}y}
-\sigma_{q_{1}x}\sigma_{\bar {q}_{2}x})
[4p'_{q_{1}x}p_{q_{1}y}+4p'_{q_{1}x}p_{\bar {q}_{2}y} 
\nonumber     \\
&-&8p'_{q_{1}x}p'_{q_{1}y}+4p'_{q_{1}y}p_{q_{1}x}+4p'_{q_{1}y}p_{\bar {q}_{2}x}
-4p_{q_{1}x}p_{q_{1}y} -4p_{\bar {q}_{2}y}p_{\bar {q}_{2}x}]
+i\sigma_{q_{1}z}\sigma_{\bar {q}_{2}y}[p_{q_{1}z}p_{\bar {q}_{2}x}
\nonumber     \\
&+&3p'_{q_{1}z}p_{q_{1}x} +3p'_{q_{1}z}p_{\bar
{q}_{2}x}-4p'_{q_{1}z}p'_{q_{1}x}+p'_{q_{1}x}p_{\bar {q}_{2}z}
+p'_{q_{1}x}p_{q_{1}z}-p_{q_{1}x}p_{\bar {q}_{2}z}
\nonumber     \\
&-&2p_{\bar {q}_{2}x}p_{\bar {q}_{2}z}-2p_{q_{1}x}p_{q_{1}z}]
+i\sigma_{q_{1}z}\sigma_{\bar {q}_{2}x}[-p_{q_{1}z}p_{\bar
{q}_{2}z}-3p'_{q_{1}z}p_{q_{1}y}-3p'_{q_{1}z}p_{\bar{q}_{2}y}
\nonumber     \\
&+&4p'_{q_{1}z}p'_{q_{1}y}-p'_{q_{1}y}p_{\bar {q}_{2}z}
-p'_{q_{1}y}p_{q_{1}z}+2p_{q_{1}z}p_{q_{1}y}+2p_{\bar
{q}_{2}y}p_{\bar {q}_{2}z}+p_{q_{1}y}p_{\bar {q}_{2}z}]\} .
\nonumber     \\
\end{eqnarray}
Replacing the subscripts, $x$ in Eq. (20) with $y$ ($z$), $y$ with $z$ 
($x$), and $z$ with $x$ ($y$), we get $[S_x,H_1]$ ($[S_y,H_1]$). Obviously,
$[S_x,H_1]$, $[S_y,H_1]$, and $[S_z,H_1]$ may not be zero. Therefore, $S_x$, 
$S_y$, and $S_z$ may not be conserved, and the total spin may not be conserved,
i.e., the total spin of the two final mesons
may not equal the total spin of the two initial mesons.

For the right diagram in Fig.~\ref{fig1}, the quark-antiquark
potential corresponding to 
$\bar q_1+q_2 \to q_3 + \bar q_4$ is between $q_2$ and $\bar{q}_1$,
\begin{eqnarray}
V_{{\rm a}q_2\bar{q}_1\rm F}(\vec{k})&=&
-\frac{g^{2}_s}{k^2}\left[\frac{1}{3}\lambda^{0}_{f}(32)\lambda^{0}_{f}(41)
+\frac{1}{2}\vec{\lambda}_{f}(32)\cdot\vec{\lambda}^{T}_{f}(41)\right]
 \nonumber     \\
&\times&\left[\frac{4}{9}\lambda^{0}(32)\lambda^{0}(41)-\frac{1}{12}
\vec{\lambda}(32)\cdot\vec{\lambda}^{T}(41)\right]
\left[-\frac{3}{2}-\frac{1}{2}\vec{\sigma}(32)\cdot\vec{\sigma}(41) \right.
\nonumber     \\
&+&\frac{\vec{\sigma}(41)\cdot\vec{p}_{{\bar q_4}}\vec{\sigma}(41)\cdot
\vec{p}_{{{\bar q_1}}}}{8m_{q_2}m_{q_3}}
+\frac{\vec{\sigma}(32)\cdot\vec{p}_{q_3}\vec{\sigma}(32)\cdot
\vec{p}_{q_2}}{8m_{q_2}m_{q_3}}
+\frac{3\vec{\sigma}(32)\cdot\vec{p}_{q_2} \vec{\sigma}(41)\cdot
\vec{p}_{\bar{q}_1}}{8m^{2}_{q_2}}
\nonumber     \\
&-&\frac{\vec{\sigma}(32)\cdot\vec{p}_{q_2} \vec{\sigma}(41)\cdot
\vec{p}_{\bar q_4}}{8m_{q_2}m_{q_3}}
-\frac{\vec{\sigma}(32)\cdot\vec{p}_{q_3} \vec{\sigma}(41)\cdot
\vec{p}_{\bar q_1}}{8m_{q_2}m_{q_3}}
+\frac{3\vec{\sigma}(32)\cdot\vec{p}_{q_3} \vec{\sigma}(41)\cdot
\vec{p}_{\bar q_4}}{8m^{2}_{q_3}}
\nonumber     \\
&+&\frac{\vec{\sigma}(32)\vec{\sigma}(32)\cdot\vec{p}_{q_2} \vec{\sigma}(41)
\vec{\sigma}(41)\cdot\vec{p}_{\bar q_1}}{8m^{2}_{q_2}}
+\frac{\vec{\sigma}(32)\vec{\sigma}(32)\cdot\vec{p}_{q_2} \vec{\sigma}(41)
\cdot\vec{p}_{\bar q_4}\vec{\sigma}(41)}{8m_{q_2}m_{q_3}}
 \nonumber     \\
&+&\frac{\vec{\sigma}(32)\cdot\vec{p}_{q_3}\vec{\sigma}(32)\cdot 
\vec{\sigma}(41)\vec{\sigma}(41)\cdot\vec{p}_{\bar q_1}}{8m_{q_2}m_{q_3}} 
+\frac{\vec{\sigma}(32)\cdot\vec{p}_{q_3}\vec{\sigma}(32)\vec{\sigma}(41)
\cdot\vec{p}_{\bar q_4}\vec{\sigma}(41)}{8m^{2}_{q_3}}
 \nonumber     \\
&-&\left. \frac{\vec{\sigma}(41)\cdot\vec{p}_{\bar q_4}\vec{\sigma}(32)\cdot 
\vec{\sigma}(41)\vec{\sigma}(41)\cdot\vec{p}_{\bar q_1}}{8m_{q_2}m_{q_3}}
-\frac{\vec{\sigma}(32)\cdot\vec{p}_{q_3}\vec{\sigma}(32)\cdot 
\vec{\sigma}(41)\vec{\sigma}(32)\cdot\vec{p}_{q_2}}{8m_{q_2}m_{q_3}}\right] .
 \nonumber     \\
\end{eqnarray}
Required by the Fierz transformations, quark 3 (antiquark 4) and quark 2
(antiquark 1) have the same flavor, and the quark and antiquark masses
satisfy $m_{q_3}=m_{q_2}$ and $m_{\bar{q}_4}=m_{\bar{q}_1}$.
The Hamiltonian corresponding to the right diagram in Fig.~\ref{fig1} is
\begin{eqnarray}
H_2 &=& V_{{\rm a}q_2 \bar{q}_1\rm F}(\vec{k})+V_{q_1\bar{q}_1}
+V_{q_2\bar{q}_2}
\nonumber     \\
&+&\sqrt {m^2_{q_1}+\vec{p}^{~2}_{q_1}}
+\sqrt {m^2_{\bar{q}_1}+\vec{p}^{~2}_{\bar{q}_1}}
+\sqrt {m^2_{q_2}+\vec{p}^{~2}_{q_2}}
+\sqrt {m^2_{\bar{q}_2}+\vec{p}^{~2}_{\bar{q}_2}} .
\end{eqnarray}
We may obtain the commutator $[S_z,H_2]$ from $[S_z,H_1]$ in Eq. (20) by the
replacement, $q_1 \leftrightarrow q_2$ and 
$\bar{q}_1 \leftrightarrow \bar{q}_2$. We may also 
get $[S_x,H_2]$ and $[S_y,H_2]$ from $[S_x,H_1]$ and $[S_y,H_1]$ by the
replacement. $[S_x,H_2]$, $[S_y,H_2]$, and $[S_z,H_2]$ may not be zero. This
indicates that $S_x$, $S_y$, and $S_z$ may not be conserved, and the total spin
may not be conserved, i.e., the total spin of the two final mesons
may not equal the total spin of the two initial mesons.

The development in spherical harmonics of the relative-motion wave function
of mesons $A$ and $B$ (aside from a normalization constant) is given by
\begin{eqnarray}
e^{i\vec{p}_{q_1\bar{q}_1,q_2\bar{q}_2} \cdot 
\vec{r}_{q_1\bar{q}_1,q_2\bar{q}_2}} & = & 4\pi 
\sum\limits_{L_{\rm i}=0}^{\infty} 
\sum\limits_{M_{\rm i}=-L_{\rm i}}^{L_{\rm i}}
i^{L_{\rm i}} j_{L_{\rm i}} (\mid \vec{p}_{q_1\bar{q}_1,q_2\bar{q}_2} \mid
r_{q_1\bar{q}_1,q_2\bar{q}_2}) 
\nonumber     \\
& \times & Y_{L_{\rm i}M_{\rm i}}^\ast
(\hat{p}_{q_1\bar{q}_1,q_2\bar{q}_2}) Y_{L_{\rm i}M_{\rm i}}
(\hat{r}_{q_1\bar{q}_1,q_2\bar{q}_2}).
\end{eqnarray}
The development of the relative-motion wave function of mesons $C$ and $D$
(aside from a normalization constant) is
\begin{eqnarray}
e^{i\vec{p}_{q_3\bar{q}_1,q_2\bar{q}_4} \cdot 
\vec{r}_{q_3\bar{q}_1,q_2\bar{q}_4}} & = & 4\pi 
\sum\limits_{L_{\rm f}=0}^{\infty} 
\sum\limits_{M_{\rm f}=-L_{\rm f}}^{L_{\rm f}} 
i^{L_{\rm f}} j_{L_{\rm f}} (\mid \vec{p}_{q_3\bar{q}_1,q_2\bar{q}_4} \mid
r_{q_3\bar{q}_1,q_2\bar{q}_4}) 
\nonumber     \\
& \times & Y_{L_{\rm f}M_{\rm f}}^\ast
(\hat{p}_{q_3\bar{q}_1,q_2\bar{q}_4}) Y_{L_{\rm f}M_{\rm f}}
(\hat{r}_{q_3\bar{q}_1,q_2\bar{q}_4}),
\end{eqnarray}
in ${\cal M}_{{\rm a}q_1\bar {q}_2}$, and 
\begin{eqnarray}
e^{i\vec{p}_{q_1\bar{q}_4,q_3\bar{q}_2} \cdot 
\vec{r}_{q_1\bar{q}_4,q_3\bar{q}_2}} & = & 4\pi 
\sum\limits_{L_{\rm f}=0}^{\infty} 
\sum\limits_{M_{\rm f}=-L_{\rm f}}^{L_{\rm f}} 
i^{L_{\rm f}} j_{L_{\rm f}} (\mid \vec{p}_{q_1\bar{q}_4,q_3\bar{q}_2} \mid
r_{q_1\bar{q}_4,q_3\bar{q}_2}) 
\nonumber     \\
& \times & Y_{L_{\rm f}M_{\rm f}}^\ast
(\hat{p}_{q_1\bar{q}_4,q_3\bar{q}_2}) Y_{L_{\rm f}M_{\rm f}}
(\hat{r}_{q_1\bar{q}_4,q_3\bar{q}_2}),
\end{eqnarray}
in ${\cal M}_{{\rm a}\bar{q}_1q_2}$. $Y_{LM}(\hat{r})$ are the spherical 
harmonics with the orbital-angular-momentum quantum number $L$ and the magnetic
quantum number $M$, and $\hat{r}$ denotes the polar angles of $\vec{r}$.
Let $S$ ($S^\prime$) and $S_z$ ($S_z^\prime$) be the total spin of mesons $A$
($C$) and $B$ ($D$) and its $z$ component, respectively. In the transition
amplitudes we have
\begin{eqnarray}
Y_{L_{\rm i}M_{\rm i}}\chi_{S_AS_{Az}}\chi_{S_BS_{Bz}} & = &
\sum\limits_{S=\mid S_A-S_B \mid}^{S_A+S_B}
\sum\limits_{S_z=-S}^{S} (S_AS_{Az}S_BS_{Bz} \mid SS_z)
\nonumber     \\
& \times & \sum\limits_{J=\mid L_{\rm i}-S \mid}^{L_{\rm i}+S}
\sum\limits_{J_z=-J}^{J}
(L_{\rm i}M_{\rm i}SS_z \mid JJ_z) \varphi_{JJ_z}^{\rm i},
\end{eqnarray}
\begin{eqnarray}
Y_{L_{\rm f}M_{\rm f}}\chi_{S_CS_{Cz}}\chi_{S_DS_{Dz}} & = &
\sum\limits_{S^\prime =\mid S_C-S_D \mid}^{S_C+S_D}
\sum\limits_{S_z^\prime =-S^\prime}^{S^\prime}
(S_CS_{Cz}S_DS_{Dz} \mid S^\prime S_z^\prime)
\nonumber     \\
& \times & \sum\limits_{J^\prime=\mid L_{\rm f}
-S^\prime \mid}^{L_{\rm f}+S^\prime}
\sum\limits_{J_z^\prime=-J^\prime}^{J^\prime}
(L_{\rm f}M_{\rm f}S^\prime S_z^\prime 
\mid J^\prime J_z^\prime) \varphi_{J^\prime J_z^\prime}^{\rm f},
\end{eqnarray}
where the Clebsch-Gordan coefficients are used.
The spherical harmonics and the spin wave functions are coupled to
the wave functions $\varphi_{JJ_z}^{\rm i}$ and 
$\varphi_{J^\prime J_z^\prime}^{\rm f}$, where
$J$ ($J^\prime$) is the total-angular-momentum quantum number of the 
two initial (final) mesons, and $J_z$ ($J_z^\prime$) is its $z$ component.
Even though the total spin of the two final mesons may not equal the total
spin of the two initial mesons, in the reactions listed in Table 1 the total 
angular momentum of the two final mesons equals the total
angular momentum of the two initial mesons. In addition, parity
is conserved. The parity conservation connects the orbital-angular-momentum
quantum numbers in the relative-motion wave functions of the initial and final
mesons. The orbital-angular-momentum quantum numbers are selected to
satisfy the parity conservation, symmetrization of wave functions of identical
bosons, and $J=J^\prime$. For example, $L_{\rm i}=L_{\rm f}=J$ excluding 
$L_{\rm i}=L_{\rm f}=0$ is required in $K\bar{K} \to K\bar{K}^*$, 
$K\bar{K} \to K^*\bar{K}$, $\pi K \to \pi K^*$,
$\pi K \to \rho K$, and $\pi \rho \to K\bar{K}$; $L_{\rm i}=L_{\rm f}=J$ with 
$L_{\rm i}>0$, odd $L_{\rm i}$ for $I=1$, and even $L_{\rm i}$ for $I=0$ in 
$\pi \pi \to K\bar{K}^*$ and $\pi \pi \to K^* \bar{K}$. In practical 
calculations the summations over $L_{\rm i}$ in Eq. (23) and over $L_{\rm f}$
in Eqs. (24) and (25) are from 0 to 3.

\vspace{0.5cm}
\leftline{\bf V. SUMMARY }
\vspace{0.5cm}

From the transition potential and the mesonic quark-antiquark relative-motion
wave functions, we have calculated the unpolarized cross sections for the 
2-to-2 meson-meson reactions that arise from quark-antiquark annihilation and 
creation in the first Born approximation. The reactions include
$K \bar {K} \to K \bar {K}^\ast $,~$ K\bar{K} \to K^\ast \bar{K}$,
$\pi K \to \pi K^\ast$,~$\pi K \to \rho K$,~$\pi \pi \to K \bar{K}^\ast$,
$\pi \pi \to K^* \bar K$,~$\pi \pi \to K^\ast\bar{K}^\ast$,
$\pi \rho \to K \bar{K}$,~$\pi \rho \to K^\ast\bar{K}^\ast$,
$\rho \rho \to K^\ast\bar{K}^\ast$, $K \bar{K}^\ast \to \rho \rho$, and
$K^* \bar{K} \to \rho \rho$.
The Hamiltonian of the two mesons contains the quark-antiquark potential which
is equivalent to the transition potential. We have derived the commutation 
relations of the total spin of the two mesons and the Hamiltonian. Due to the 
quark-antiquark potential the commutators may not be zero, 
and the total spin may 
not be conserved in the reactions. With increasing center-of-mass energy of
the two initial mesons from the threshold energy, the
cross sections for the endothermic reactions increase very rapidly to the peak
cross sections first, and then decrease or display plateaus for 
$\pi \pi \to K^\ast\bar{K}^\ast$ and $\pi \rho \to K^\ast\bar{K}^\ast$
at some temperatures. The cross sections exhibit remarkable temperature
dependence. To use the cross sections conveniently,
we have parametrized the numerical cross sections for the ten isospin channels 
of reactions. Based on the flavor matrix elements, the cross sections for the
other isospin channels of reactions can be obtained from the cross sections
for the ten channels.

\vspace{0.5cm}
\leftline{\bf ACKNOWLEDGEMENTS}
\vspace{0.5cm}

This work was supported by the National Natural Science Foundation
of China under Grant No. 11175111.

\newpage

\newpage
\begin{figure}[htbp]
  \centering
    \includegraphics[scale=0.85]{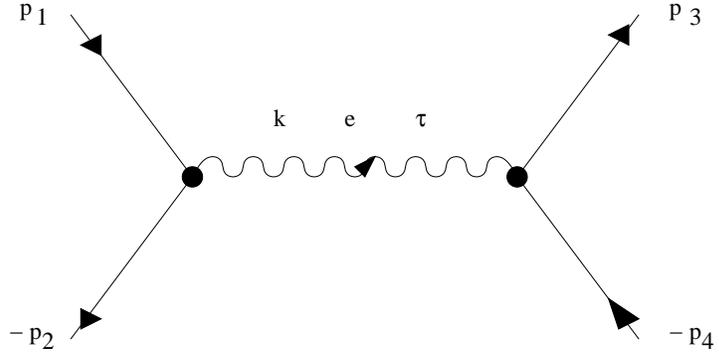}
\caption{Corresponding to the process $q(p_1)+\bar{q}(-p_2) \to q'(p_3)
+\bar{q}'(-p_4)$, solid (wavy) lines stand for quarks or antiquarks (gluon).
$k$ denotes the gluon four-momentum, $e$ its color index, and $\tau$ its
space-time index.}
\label{fig2}
\end{figure}

\newpage
\begin{figure}[htbp]
  \centering
    \includegraphics[scale=0.65]{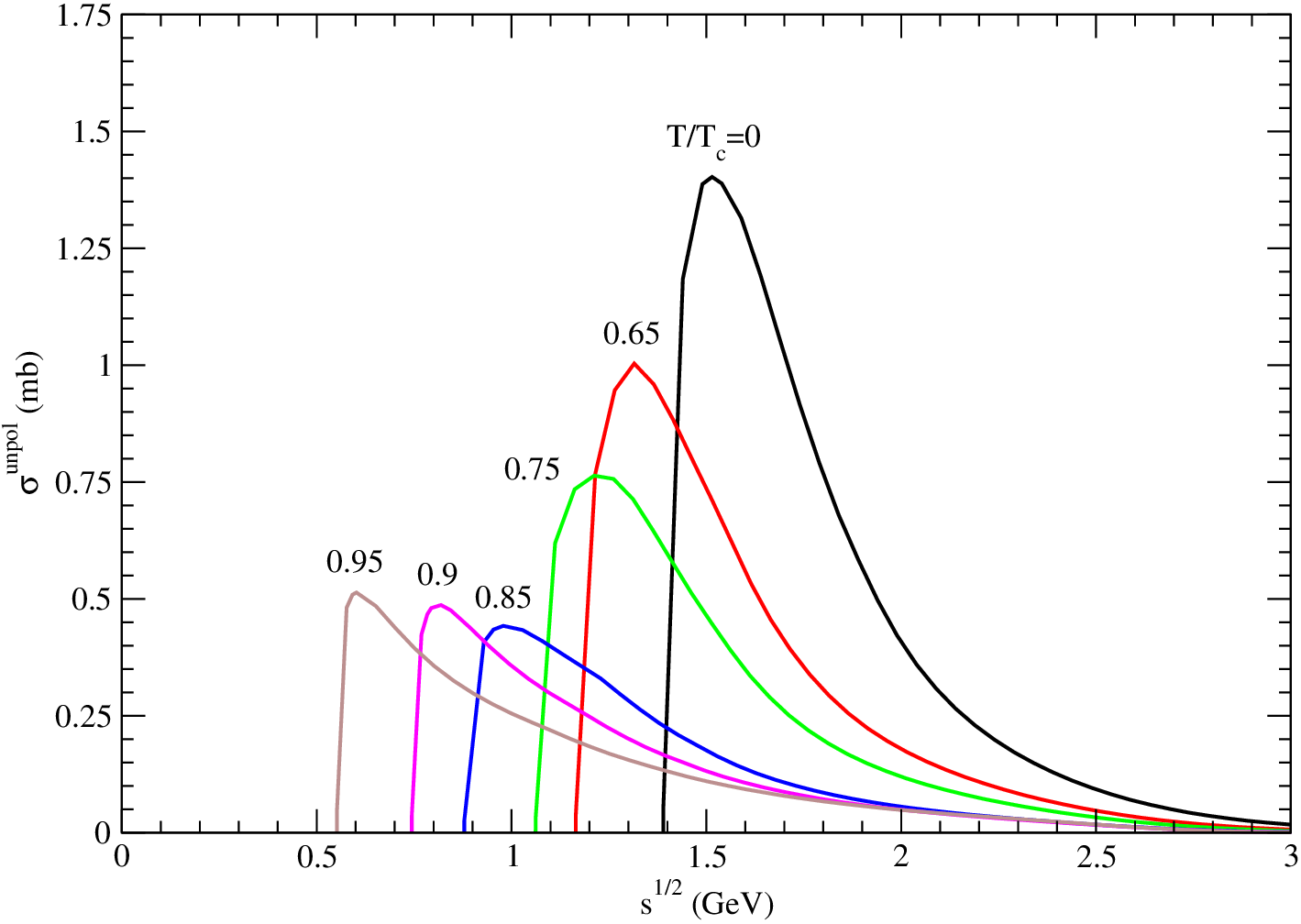}
\caption{Cross sections for $K \bar{K} \to K \bar{K}^\ast$ for $I=1$
at various temperatures.}
\label{fig3}
\end{figure}

\newpage
\begin{figure}[htbp]
  \centering
    \includegraphics[scale=0.65]{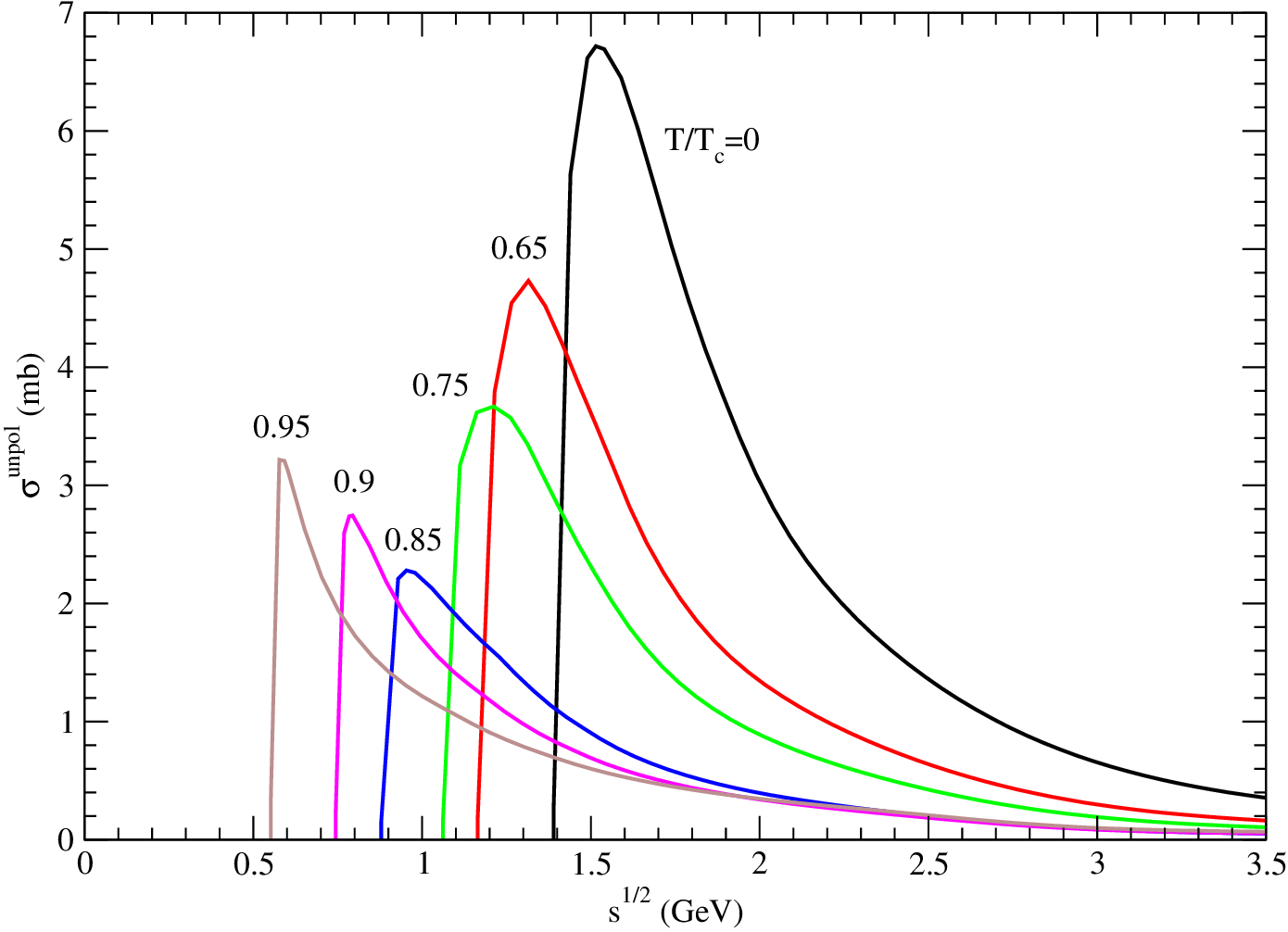}
\caption{Cross sections for $K \bar{K} \to K \bar{K}^\ast$ for $I=0$
at various temperatures.}
\label{fig4}
\end{figure}

\newpage
\begin{figure}[htbp]
  \centering
    \includegraphics[scale=0.65]{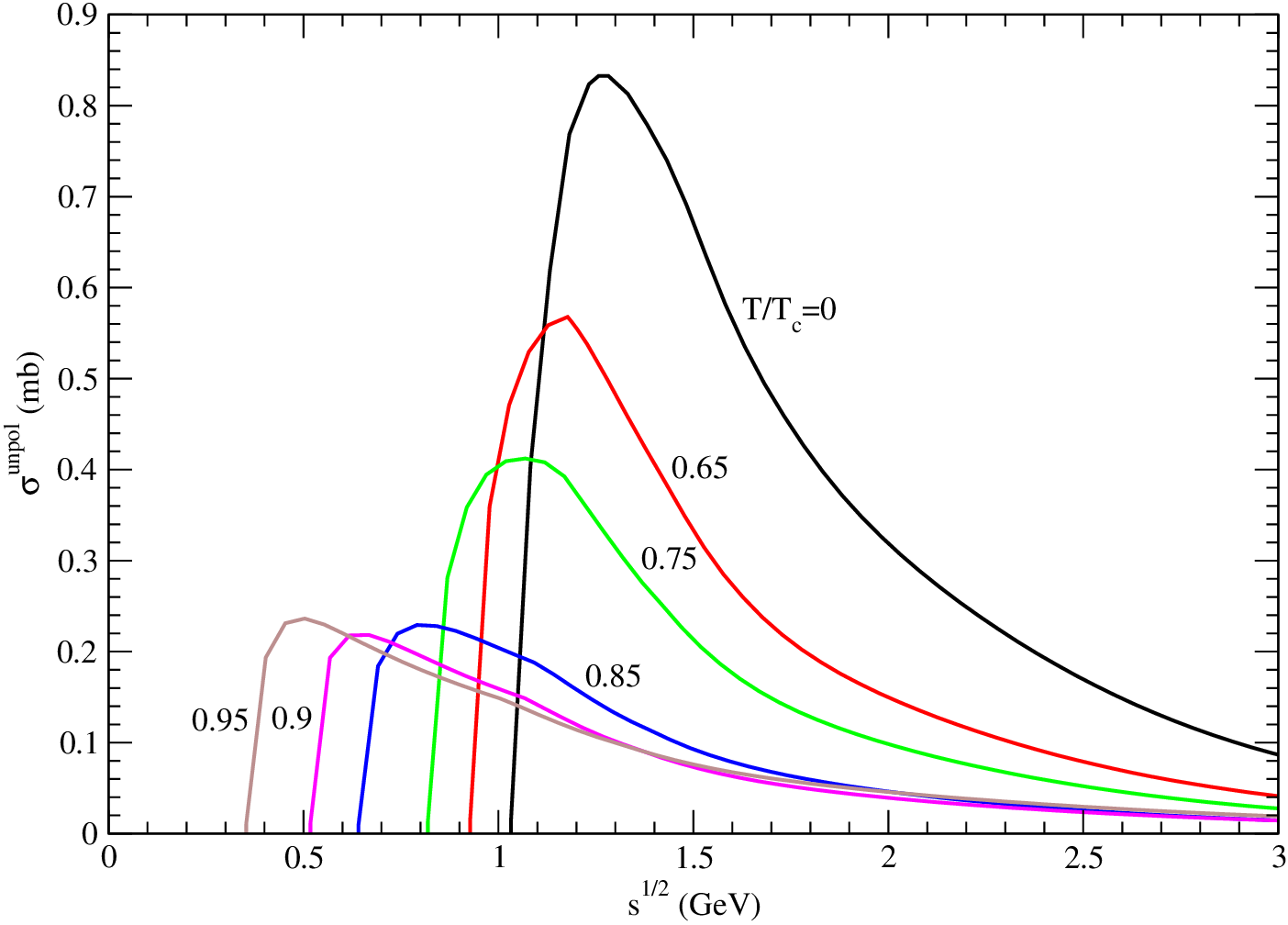}
\caption{Cross sections for $\pi K \to \pi K^\ast$ for $I=1/2$
at various temperatures.}
\label{fig5}
\end{figure}

\newpage
\begin{figure}[htbp]
  \centering
    \includegraphics[scale=0.65]{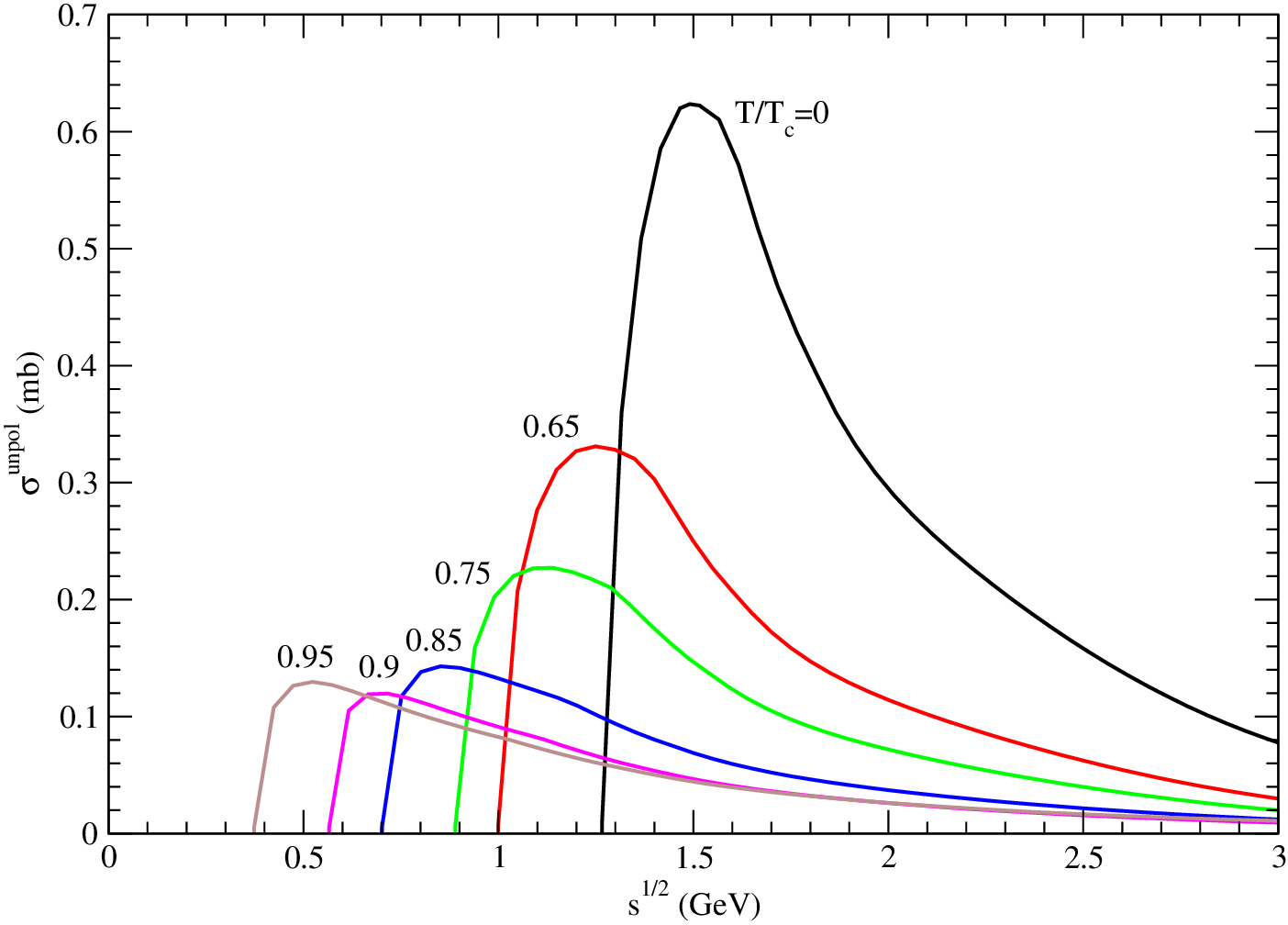}
\caption{Cross sections for $\pi K \to \rho K$ for $I=1/2$
at various temperatures.}
\label{fig6}
\end{figure}

\newpage
\begin{figure}[htbp]
  \centering
    \includegraphics[scale=0.65]{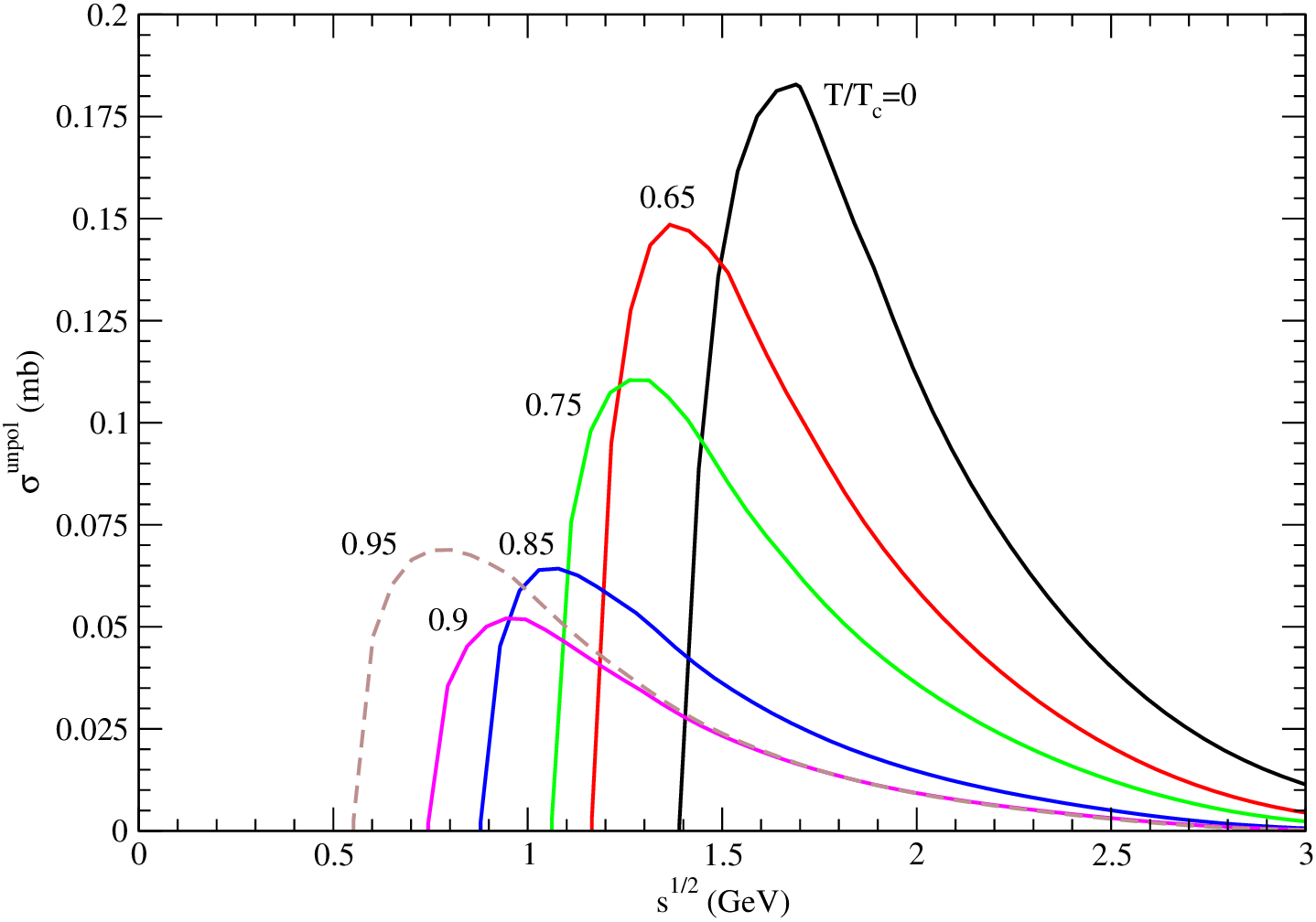}
\caption{Cross sections for $\pi \pi \to K \bar{K}^\ast$ for $I=1$
at various temperatures.}
\label{fig7}
\end{figure}

\newpage
\begin{figure}[htbp]
  \centering
    \includegraphics[scale=0.65]{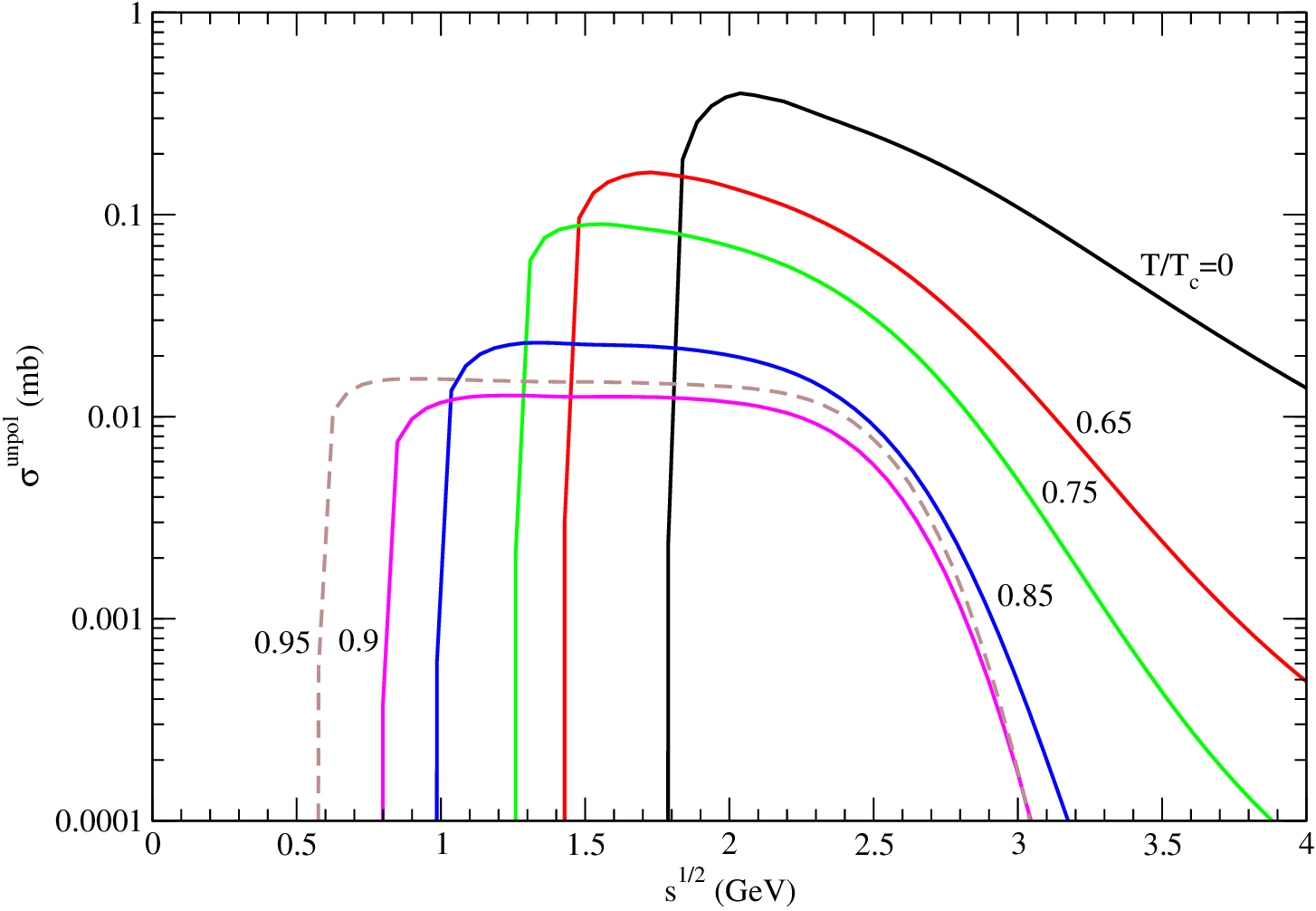}
\caption{Cross sections for $\pi \pi \to K^\ast \bar{K}^\ast$ for $I=1$
at various temperatures.}
\label{fig8}
\end{figure}

\newpage
\begin{figure}[htbp]
  \centering
    \includegraphics[scale=0.65]{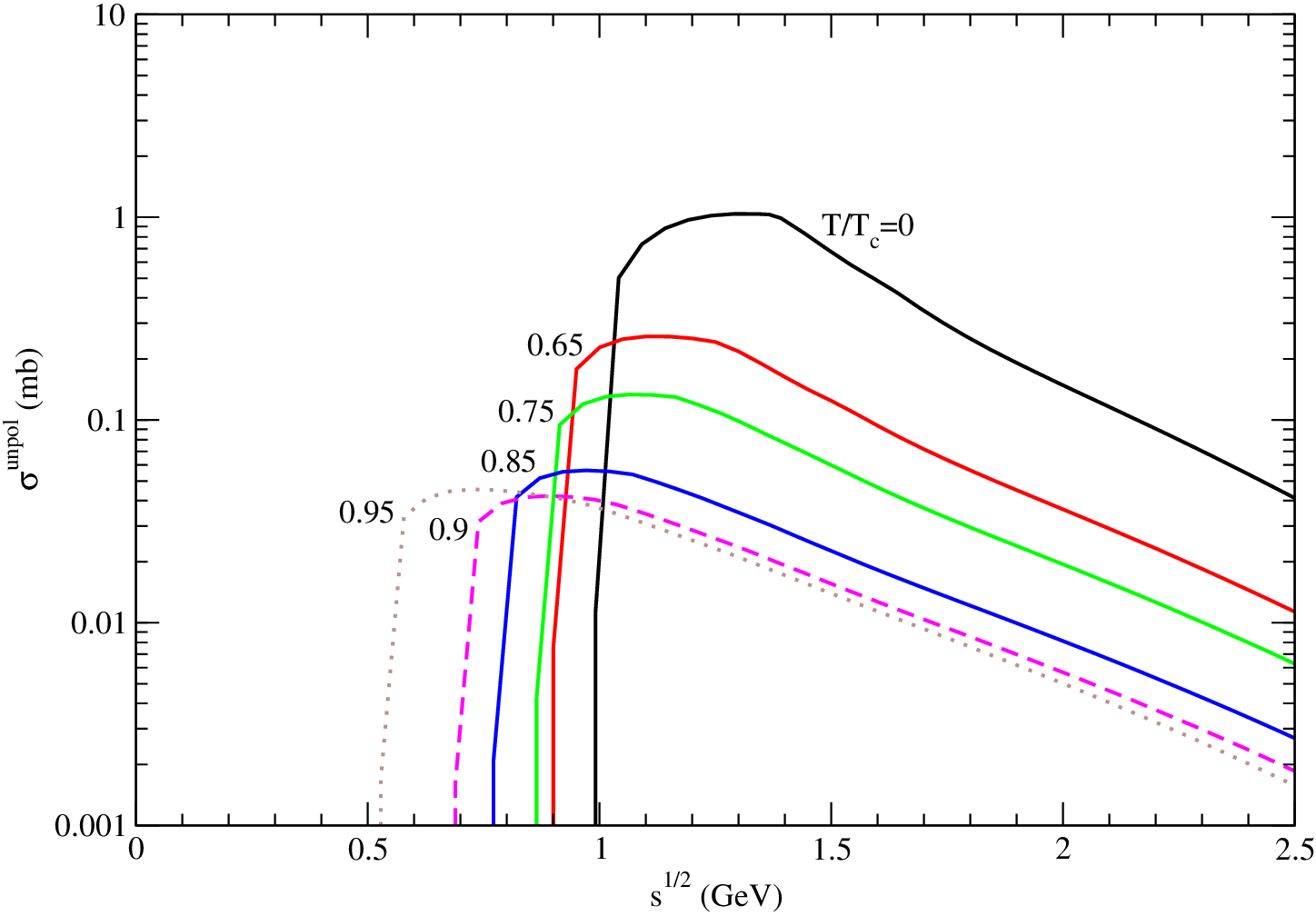}
\caption{Cross sections for $\pi \rho \to K \bar{K}$ for $I=1$
at various temperatures.}
\label{fig9}
\end{figure}

\newpage
\begin{figure}[htbp]
  \centering
    \includegraphics[scale=0.65]{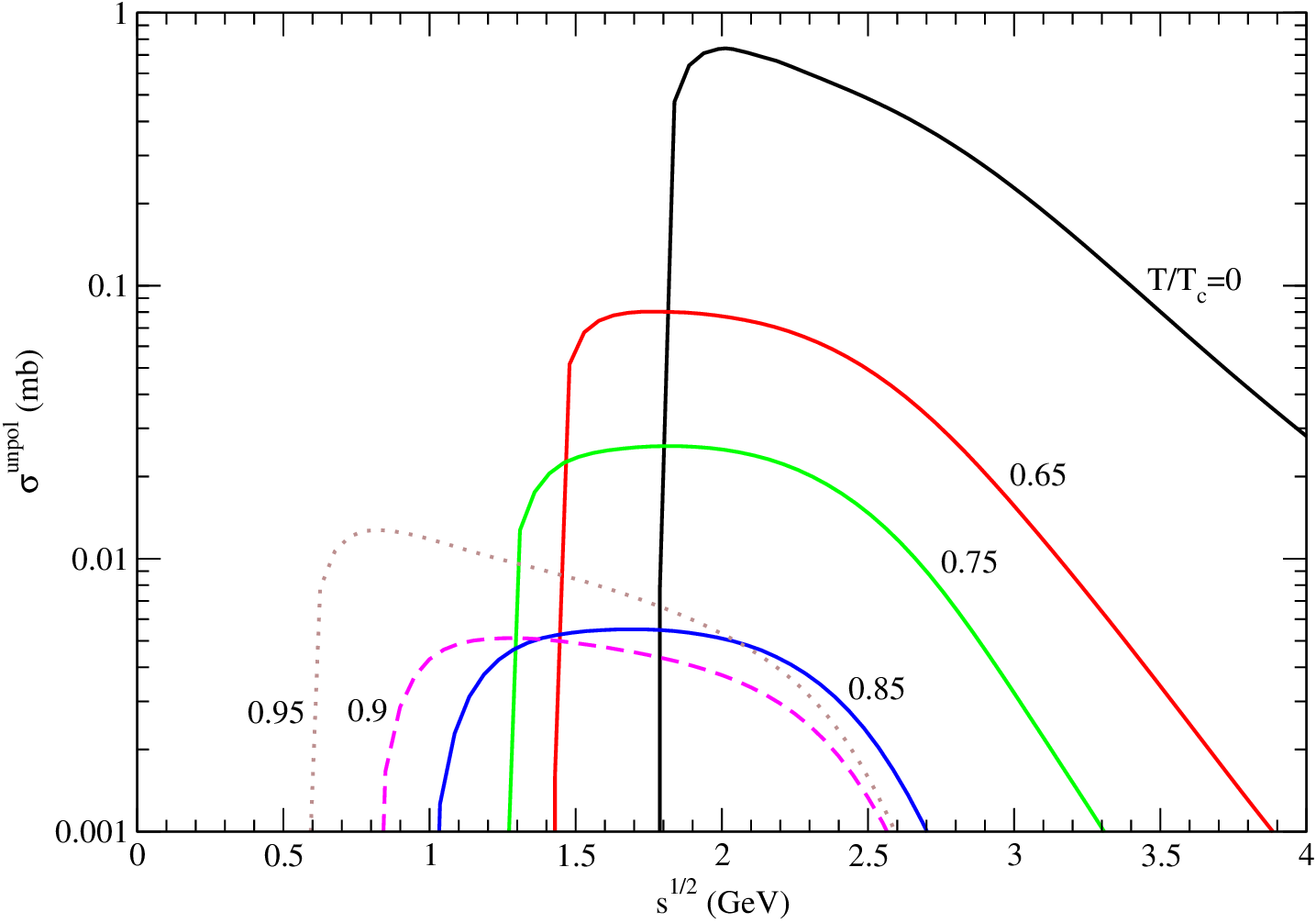}
\caption{Cross sections for $\pi \rho \to K^\ast \bar{K}^\ast$ for $I=1$
at various temperatures.}
\label{fig10}
\end{figure}

\newpage
\begin{figure}[htbp]
  \centering
    \includegraphics[scale=0.65]{rrkaka_a1.eps}
\caption{Cross sections for $\rho \rho \to K^\ast \bar{K}^\ast$ for $I=1$
at various temperatures.}
\label{fig11}
\end{figure}

\newpage
\begin{figure}[htbp]
  \centering
    \includegraphics[scale=0.65]{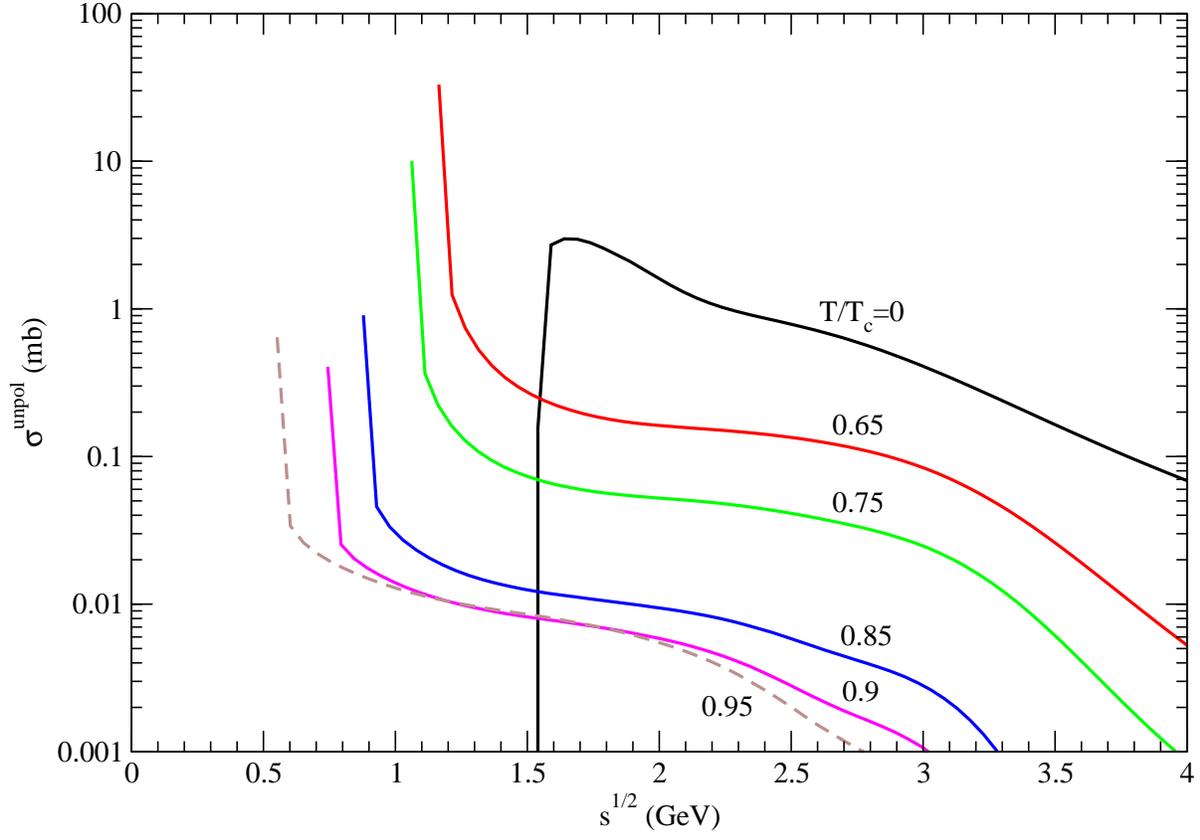}
\caption{Cross sections for $K \bar{K}^\ast \to \rho \rho$ for $I=1$
at various temperatures.}
\label{fig12}
\end{figure}

\newpage
\begin{table*}[htbp]
\caption{\label{table1} Flavor matrix elements.}
\begin{tabular}{ccc}
\hline
Channel & ${\cal M}_{{\rm a}q_1\bar{q}_2\rm f}$ &
${\cal M}_{{\rm a}\bar{q}_1 q_2\rm f}$ \\
\hline
$I=1~ K \bar {K} \to K \bar {K}^\ast$ & 0 & 1  \\
$I=0~ K \bar {K} \to K \bar {K}^\ast$ & 2 & 1  \\
$I=1~ K \bar {K} \to K^* \bar {K}$ & 0 & 1  \\
$I=0~ K \bar {K} \to K^* \bar {K}$ & 2 & 1  \\
$I=3/2~ \pi K \to \pi K^\ast$ & 0 & 0  \\
$I=1/2~ \pi K \to \pi K^\ast$ & 0 & $\frac{3}{2}$  \\
$I=3/2~ \pi K \to \rho K$ & 0 & 0  \\
$I=1/2~ \pi K \to \rho K$ & 0 & $\frac{3}{2}$  \\
$I=1~ \pi \pi \to K \bar{K}^\ast$ & 0 & -1  \\
$I=0~ \pi \pi \to K \bar{K}^\ast$ & 0 & -$\frac {\sqrt 6}{2}$  \\
$I=1~ \pi \pi \to K^* \bar{K}$ & 0 & -1  \\
$I=0~ \pi \pi \to K^* \bar{K}$ & 0 & -$\frac {\sqrt 6}{2}$  \\
$I=1~ \pi \pi \to K^* \bar{K}^*$ & 0 & -1  \\
$I=0~ \pi \pi \to K^* \bar{K}^*$ & 0 & -$\frac {\sqrt 6}{2}$  \\
$I=1~ \pi \rho \to K \bar {K}$ & 0 & -1  \\
$I=0~ \pi \rho \to K \bar {K}$ & 0 & -$\frac {\sqrt 6}{2}$  \\
$I=1~ \pi \rho \to K^* \bar {K}^*$ & 0 & -1  \\
$I=0~ \pi \rho \to K^* \bar {K}^*$ & 0 & -$\frac {\sqrt 6}{2}$  \\
$I=1~ \rho \rho \to K^\ast \bar {K}^*$ & 0 & -1  \\
$I=0~ \rho \rho \to K^\ast \bar {K}^*$ & 0 & -$\frac {\sqrt 6}{2}$  \\
$I=1~ K \bar {K}^* \to \rho \rho$ & 0 & -1  \\
$I=0~ K \bar {K}^* \to \rho \rho$ & 0 & -$\frac {\sqrt 6}{2}$  \\
$I=1~ K^* \bar {K} \to \rho \rho$ & 0 & -1  \\
$I=0~ K^* \bar {K} \to \rho \rho$ & 0 & -$\frac {\sqrt 6}{2}$  \\
\hline
\end{tabular}
\end{table*}

\newpage
\begin{table*}[htbp]
\caption{\label{table2}Values of the parameters. $a_1$ and $a_2$ are
in units of millibarns; $b_1$, $b_2$, $d_0$, and $\sqrt{s_{\rm z}}$ are
in units of GeV; $e_1$ and $e_2$ are dimensionless.}
\tabcolsep=5pt
\begin{tabular}{cccccccccc}
  \hline
  \hline
Reactions & $T/T_{\rm c} $ & $a_1$ & $b_1$ & $e_1$ & $a_2$ & $b_2$ & $e_2$ &
$d_0$ & $\sqrt{s_{\rm z}} $\\
\hline
 $I=1~K\bar{K}\to K \bar{K}^{*}$
  &  0     & 0.14  & 0.173  & 0.3   & 1.27  & 0.124  & 0.6  & 0.125 & 3.06\\
  &  0.65  & 0.1   & 0.209  & 0.2   & 0.9   & 0.142  & 0.7  & 0.15 & 2.91\\
  &  0.75  & 0.1   & 0.26   & 1.0   & 0.69  & 0.128  & 0.5  & 0.15 & 2.87\\
  &  0.85  & 0.1   & 0.03   & 0.5   & 0.4 & 0.157  & 0.51   & 0.1  & 2.82\\
  &  0.9   & 0.2   & 0.04   & 0.6   & 0.35& 0.153   & 0.41  & 0.075  & 2.77\\
  &  0.95  & 0.3   & 0.21    & 0.5  & 0.32 & 0.035  & 0.45  & 0.05 & 2.76\\
  \hline
 $I=0~K\bar{K}\to K \bar{K}^{*}$
  &  0     & 4.0  & 0.108 & 0.63 & 3.05  & 0.251  & 0.44  & 0.125 & 7.21\\
  &  0.65  & 3.3  & 0.127 & 0.67 & 1.5  & 0.267  & 0.35  & 0.15 & 6.12\\
  &  0.75  & 2.5  & 0.114 & 0.56 & 1.3 & 0.25  & 0.39  & 0.15  & 5.78\\
  &  0.85  & 1.1  & 0.059 & 0.5  & 1.4 & 0.195   & 0.45  & 0.075  & 5.58\\
  &  0.9   & 1.4  & 0.183 & 0.4  & 1.7 & 0.04  & 0.5   & 0.05  & 5.34\\
  &  0.95  & 1.6  & 0.16  & 0.33 & 2.0 & 0.03 & 0.52 & 0.025 & 5.74\\
  \hline
 $ I=1/2~\pi K\to\pi K^{*} $
  &  0     & 0.4  & 0.22  & 1.5 & 0.45  & 0.329 & 0.5 & 0.225  & 6.47\\
  &  0.65  & 0.3   & 0.18 & 1.0 & 0.26 & 0.275 & 0.4 & 0.25   & 5.78\\
  &  0.75  & 0.1 & 0.33  & 0.3 & 0.33 & 0.2  & 0.8 & 0.25    & 5.71\\
  &  0.85  & 0.2 & 0.17 & 0.5 & 0.04 & 0.439 & 0.4 & 0.15  & 5.81\\
  &  0.9   & 0.07 & 0.1 & 0.9 & 0.17 & 0.219 & 0.4 & 0.15  & 6.05\\
  &  0.95  & 0.1 & 0.1 & 0.5 & 0.15 & 0.308 & 0.47  & 0.15 & 6.23\\
  \hline
 $ I=1/2~\pi K\to \rho K $
  &  0     & 0.309 & 0.33   & 0.5  & 0.33 & 0.18 & 1.1  & 0.225    & 6.22\\
  &  0.65  & 0.061 & 0.84   & 0.29 & 0.29 & 0.21  & 0.8 & 0.25   &  4.81\\
  &  0.75  & 0.15  & 0.19   & 0.8  & 0.09 & 0.34 & 0.4 & 0.25  &  4.78\\
  &  0.85  & 0.055 & 0.4    & 0.5  & 0.1 & 0.14   & 0.5 & 0.15 &  4.94\\
  &  0.9   & 0.05  & 0.61   & 1.4  & 0.11 & 0.1  & 0.5 & 0.15 &  5.02\\
  &  0.95  & 0.05  & 0.71   & 1.6  & 0.12 & 0.12 & 0.5 & 0.15  &  5.37\\
  \hline
  \hline
\end{tabular}
\end{table*}

\newpage
\begin{table*}[htbp]
\caption{\label{table3}The same as Table 2, but for three other reactions.}
\tabcolsep=5.3pt
\begin{tabular}{ccccccccccc}
  \hline
  \hline
  Reactions & $T/T_{\rm c} $ & $a_1$ & $b_1$ & $e_1$ & $a_2$ & $b_2$ & $e_2$ &
  $d_0$  & $\sqrt{s_{\rm z}} $\\
  \hline
  $I=1~\pi\pi\to K \bar{K}^{*} $
  &  0     & 0.15   & 0.25  & 0.7 & 0.03  & 0.28 & 2.2  & 0.3    & 3.69\\
  &  0.65  & 0.06   & 0.2   & 1.0 & 0.09  & 0.24  & 0.5 & 0.2   & 3.3\\
  &  0.75  & 0.016  & 0.35  & 0.4 & 0.095  & 0.19 & 0.6  & 0.2  & 3.17\\
  &  0.85  & 0.004  & 0.41  & 0.2 & 0.06 & 0.2   & 0.6  & 0.2  & 3.0\\
  &  0.9   & 0.011  & 0.26  & 0.3 & 0.041  & 0.2  & 0.7  & 0.2  & 2.91\\
  &  0.95  & 0.01  & 0.17   & 0.3 & 0.06  & 0.22  & 0.6 & 0.25  & 2.82\\
  \hline
  $I=1~\pi\pi\to K^{*} \bar{K}^{*} $
   & 0     & 0.09 & 0.22   & 1.0  & 0.3  & 0.31 & 0.7 & 0.25 & 4.76\\
   & 0.65  & 0.13 & 0.5   & 1.9  & 0.098 & 0.09 & 0.6 & 0.3 & 3.61\\
   & 0.75  & 0.024 & 0.83 & 8.0  & 0.09  & 0.225 & 0.56 & 0.3 & 3.35\\
   & 0.85  & 0.013 & 0.98 & 8.6  & 0.022 & 0.24 & 0.54 & 0.35 & 3.08\\
   & 0.9   & 0.008 & 1.13 &7.42  & 0.012 & 0.26 & 0.5 & 0.45 & 3.03\\
   & 0.95  & 0.003 &1.42 & 15.0  & 0.0181 & 0.41 & 0.5 & 0.35 & 3.01\\
   \hline
  $I=1~\pi\rho\to K \bar{K} $
   & 0     & 0.55   & 0.3 & 2.6 & 0.58& 0.174 & 0.6 & 0.3 & 2.98\\
   & 0.65  & 0.082 & 0.25 & 1.7 & 0.19 & 0.163 & 0.5 & 0.2 & 3.04\\
   & 0.75  & 0.027 & 0.22 & 1.5 & 0.11 & 0.17 & 0.5 & 0.2 & 3.09\\
   & 0.85  & 0.003 & 0.23 & 2.4 & 0.053 & 0.18 & 0.5 & 0.2 & 3.11\\
   & 0.9   & 0.005  & 0.5 & 1.3 & 0.04 & 0.16 & 0.5 & 0.2 & 3.08\\
   & 0.95  & 0.01  & 0.46 & 1.2 & 0.04 & 0.16 & 0.5 & 0.2 & 2.97\\
  \hline
  \hline
\end{tabular}
\end{table*}

\newpage
\begin{table*}[htbp]
\caption{\label{table4}The same as Table 2, but for three other reactions.}
\begin{tabular}{cccccccccc}
  \hline
  \hline
  Reactions & $T/T_{\rm c} $ & $a_1$ & $b_1$ & $e_1$ & $a_2$ & $b_2$ & $e_2$ &
  $d_0$ & $\sqrt{s_{\rm z}} $\\
  \hline
  $I=1~\pi\rho\to K^{*}\bar{K}^{*}$
  &  0     & 0.4    & 0.61 & 2.0 & 0.6  & 0.13  & 0.7 & 0.225 & 4.77\\
  &  0.65  & 0.048  & 0.78 & 3.9 & 0.073  & 0.17 & 0.6 & 0.35 & 3.96\\
  &  0.75  & 0.017   & 0.13 & 0.6 & 0.024 & 0.7  & 3.1 & 0.55 & 3.65\\
  &  0.85  & 0.0012 & 0.1 & 0.63 & 0.006 & 0.65  & 1.88 & 0.75  & 3.18\\
  &  0.9   & 0.0019 & 0.13 & 0.63 & 0.0051 & 0.66& 1.81 & 0.5  & 3.02\\
  &  0.95  & 0.01  & 0.544 & 1.1 & 0.006 & 0.1   & 0.5 & 0.25   & 2.9\\
  \hline
  $I=1~\rho\rho\to K^{*}\bar{K}^{*}$
   & 0     & 3      & 0.1   & 0.72 & 1.49    & 0.32 & 0.5  & 0.1  & 4.23\\
   & 0.65  & 0.165  & 0.16  & 0.54 & 0.1     & 0.98 & 4.5  & 0.15 & 3.98\\
   & 0.75  & 0.016  & 1.0   & 4.0  & 0.04    & 0.26 & 0.6  & 0.25 & 3.72\\
   & 0.85  & 0.0104 & 0.377 & 1.21 & 0.0017  & 0.58 & 0.57 & 0.3 & 3.25\\
   & 0.9   & 0.006  & 0.29  & 2.1  & 0.01    & 0.36 & 0.78 & 0.25  & 2.9\\
   & 0.95  & 0.017  & 0.32  & 0.5  & 0.031   & 0.21 & 1.3  & 0.2 & 2.53\\
  \hline
  $I=1~K\bar{K}^{*}\to \rho\rho$
   & 0     & 2.2    & 0.11  & 0.7  & 1.0   & 0.29 & 0.3 & 0.1  & 4.6\\
   & 0.65  & 0.18   & 0.33  & 0.23 & 0.35  & 0.05 & 0.6 & 0.05   & 4.0\\
   & 0.75  & 0.048  & 0.87  & 1.7  & 0.09  & 0.07  & 0.45 & 0.05   & 3.96\\
   & 0.85  & 0.006  & 1.18  & 4.2  & 0.012 & 0.19 & 0.51 & 0.2   & 3.56\\
   & 0.9   & 0.005  & 0.092 & 0.6  & 0.007  & 0.7 & 1.7 & 0.3   & 3.4\\
   & 0.95  & 0.002  & 1.21  & 18   & 0.01   & 0.32& 0.5 & 0.25  & 3.28\\
   \hline
   \hline
\end{tabular}
\end{table*}

\end{document}